\documentclass[longauth]{aa} 

\usepackage[switch]{lineno} 

\usepackage{pdflscape} 

\usepackage{float}

\usepackage{bm}

\usepackage{comment}

\usepackage{adjustbox}
\usepackage{array}
\usepackage{graphicx}
\usepackage{enumitem}
\usepackage{siunitx}
\usepackage{booktabs}
\usepackage[version=4]{mhchem}
\usepackage{svg}
\usepackage{soul,xcolor}
\setstcolor{red}
\usepackage{natbib}
\bibpunct{(}{)}{;}{a}{}{,} 
\usepackage{txfonts}
\usepackage[colorlinks=true,
            linkcolor=blue,
            urlcolor=blue,
            citecolor=blue]{hyperref}
\makeatletter
\renewcommand*\aa@pageof{, page \thepage{} of \pageref*{LastPage}}
\makeatother

\begin{document} 

\title{Combining \textit{Gaia} and GRAVITY: Characterising five new directly detected substellar companions}

\author{T.~O.~Winterhalder\inst{\ref{esog}}
 \and S.~Lacour\inst{\ref{lesia},\ref{esog}}
 \and A.~M\'erand\inst{\ref{esog}}
 \and J.~Kammerer\inst{\ref{esog}}
 \and A.-L.~Maire\inst{\ref{ipag}}
 \and T.~Stolker\inst{\ref{leiden}}
 \and N.~Pourr\'e\inst{\ref{ipag}}
 \and C. Babusiaux\inst{\ref{ipag}}
 \and A. Glindemann\inst{\ref{esog}}
 \and R.~Abuter\inst{\ref{esog}}
 \and A.~Amorim\inst{\ref{lisboa},\ref{centra}}
 \and R.~Asensio-Torres\inst{\ref{mpia}}
 \and W.~O.~Balmer\inst{\ref{jhupa},\ref{stsci}}
 \and M.~Benisty\inst{\ref{ipag}}
 \and J.-P.~Berger\inst{\ref{ipag}}
 \and H.~Beust\inst{\ref{ipag}}
 \and S.~Blunt\inst{\ref{northwestern}}
 \and A.~Boccaletti\inst{\ref{lesia}}
 \and M.~Bonnefoy\inst{\ref{ipag}}
 \and H.~Bonnet\inst{\ref{esog}}
 \and M.~S.~Bordoni\inst{\ref{mpe}}
 \and G.~Bourdarot\inst{\ref{mpe}}
 \and W.~Brandner\inst{\ref{mpia}}
 \and F.~Cantalloube\inst{\ref{lam}}
 \and P.~Caselli \inst{\ref{mpe}}
 \and B.~Charnay\inst{\ref{lesia}}
 \and G.~Chauvin\inst{\ref{cotedazur}}
 \and A.~Chavez\inst{\ref{northwestern}}
 \and E.~Choquet\inst{\ref{lam}}
 \and V.~Christiaens\inst{\ref{liege}}
 \and Y.~Cl\'enet\inst{\ref{lesia}}
 \and V.~Coud\'e~du~Foresto\inst{\ref{lesia}}
 \and A.~Cridland\inst{\ref{leiden}}
 \and R.~Davies\inst{\ref{mpe}}
 \and R.~Dembet\inst{\ref{lesia}}
 \and J.~Dexter\inst{\ref{boulder}}
 \and A.~Drescher\inst{\ref{mpe}}
 \and G.~Duvert\inst{\ref{ipag}}
 \and A.~Eckart\inst{\ref{cologne},\ref{bonn}}
 \and F.~Eisenhauer\inst{\ref{mpe}}
 \and N.~M.~F\"orster Schreiber\inst{\ref{mpe}}
 \and P.~Garcia\inst{\ref{centra},\ref{porto}}
 \and R.~Garcia~Lopez\inst{\ref{dublin},\ref{mpia}}
 \and T.~Gardner\inst{\ref{exeterAstro}}
 \and E.~Gendron\inst{\ref{lesia}}
 \and R.~Genzel\inst{\ref{mpe},\ref{ucb}}
 \and S.~Gillessen\inst{\ref{mpe}}
 \and J.~H.~Girard\inst{\ref{stsci}}
 \and S.~Grant\inst{\ref{mpe}}
 \and X.~Haubois\inst{\ref{esoc}}
 \and G.~Hei\ss el\inst{\ref{actesa},\ref{lesia}}
 \and Th.~Henning\inst{\ref{mpia}}
 \and S.~Hinkley\inst{\ref{exeter}}
 \and S.~Hippler\inst{\ref{mpia}}
 \and M.~Houll\'e\inst{\ref{cotedazur}}
 \and Z.~Hubert\inst{\ref{ipag}}
 \and L.~Jocou\inst{\ref{ipag}}
 \and M.~Keppler\inst{\ref{mpia}}
 \and P.~Kervella\inst{\ref{lesia}}
 \and L.~Kreidberg\inst{\ref{mpia}}
 \and N.~T.~Kurtovic\inst{\ref{mpe}}
 \and A.-M.~Lagrange\inst{\ref{ipag},\ref{lesia}}
 \and V.~Lapeyr\`ere\inst{\ref{lesia}}
 \and J.-B.~Le~Bouquin\inst{\ref{ipag}}
 \and D.~Lutz\inst{\ref{mpe}}
 \and F.~Mang\inst{\ref{mpe}}
 \and G.-D.~Marleau\inst{\ref{duisburg},\ref{tuebingen},\ref{bern},\ref{mpia}}
 \and P.~Molli\`ere\inst{\ref{mpia}}
 \and J.~D.~Monnier\inst{\ref{umich}}
 \and C.~Mordasini\inst{\ref{bern}}
 \and D.~Mouillet\inst{\ref{ipag}}
 \and E.~Nasedkin\inst{\ref{mpia}}
 \and M.~Nowak\inst{\ref{cam}}
 \and T.~Ott\inst{\ref{mpe}}
 \and G.~P.~P.~L.~Otten\inst{\ref{sinica}}
 \and C.~Paladini\inst{\ref{esoc}}
 \and T.~Paumard\inst{\ref{lesia}}
 \and K.~Perraut\inst{\ref{ipag}}
 \and G.~Perrin\inst{\ref{lesia}}
 \and O.~Pfuhl\inst{\ref{esog}}
 \and L.~Pueyo\inst{\ref{stsci}}
 \and D.~C.~Ribeiro\inst{\ref{mpe}}
 \and E.~Rickman\inst{\ref{esa}}
 \and Z.~Rustamkulov\inst{\ref{jhueps}}
 \and J.~Shangguan\inst{\ref{mpe}}
 \and T.~Shimizu \inst{\ref{mpe}}
 \and D.~Sing\inst{\ref{jhupa},\ref{jhueps}}
 \and J.~Stadler\inst{\ref{mpa},\ref{origins}}
 \and O.~Straub\inst{\ref{origins}}
 \and C.~Straubmeier\inst{\ref{cologne}}
 \and E.~Sturm\inst{\ref{mpe}}
 \and L.~J.~Tacconi\inst{\ref{mpe}}
 \and E.F.~van~Dishoeck\inst{\ref{leiden},\ref{mpe}}
 \and A.~Vigan\inst{\ref{lam}}
 \and F.~Vincent\inst{\ref{lesia}}
 \and S.~D.~von~Fellenberg\inst{\ref{bonn}}
 \and J.~J.~Wang\inst{\ref{northwestern}}
 \and F.~Widmann\inst{\ref{mpe}}
 \and J.~Woillez\inst{\ref{esog}}
 \and \c{S}.~Yaz\i{}c\i{}\inst{\ref{mpe}}
 \and  the GRAVITY Collaboration}
\institute{ 
   European Southern Observatory, Karl-Schwarzschild-Stra\ss e 2, 85748 Garching, Germany
\label{esog}      \and
   LESIA, Observatoire de Paris, PSL, CNRS, Sorbonne Universit\'e, Universit\'e de Paris, 5 place Janssen, 92195 Meudon, France
\label{lesia}      \and
   Univ.\ Grenoble Alpes, CNRS, IPAG, 38000 Grenoble, France
\label{ipag}      \and
   Universidade de Lisboa - Faculdade de Ci\^encias, Campo Grande, 1749-016 Lisboa, Portugal
\label{lisboa}      \and
   CENTRA - Centro de Astrof\' isica e Gravita\c c\~ao, IST, Universidade de Lisboa, 1049-001 Lisboa, Portugal
\label{centra}      \and
   Max-Planck-Institut f\"ur Astronomie, K\"onigstuhl 17, 69117 Heidelberg, Germany
\label{mpia}      \and
   Department of Physics \& Astronomy, Johns Hopkins University, 3400 N. Charles Street, Baltimore, MD 21218, USA
\label{jhupa}      \and
   Space Telescope Science Institute, 3700 San Martin Drive, Baltimore, MD 21218, USA
\label{stsci}      \and
   Center for Interdisciplinary Exploration and Research in Astrophysics (CIERA) and Department of Physics and Astronomy, Northwestern University, Evanston, IL 60208, USA
\label{northwestern}      \and
   Max-Planck-Institut f\"ur Extraterrestrische Physik, Giessenbachstra\ss e~1, 85748 Garching, Germany
\label{mpe}      \and
   Aix Marseille Univ, CNRS, CNES, LAM, Marseille, France
\label{lam}      \and
   Université Côte d’Azur, Observatoire de la Côte d’Azur, CNRS, Laboratoire Lagrange, France
\label{cotedazur}      \and
  STAR Institute, Universit\'e de Li\`ege, All\'ee du Six Ao\^ut 19c, 4000 Li\`ege, Belgium
\label{liege}      \and
   Leiden Observatory, Leiden University, P.O. Box 9513, 2300 RA Leiden, The Netherlands
\label{leiden}      \and
   Department of Astrophysical \& Planetary Sciences, JILA, Duane Physics Bldg., 2000 Colorado Ave, University of Colorado, Boulder, CO 80309, USA
\label{boulder}      \and
   1.\ Institute of Physics, University of Cologne, Z\"ulpicher Stra\ss e 77, 50937 Cologne, Germany
\label{cologne}      \and
   Max-Planck-Institut f\"ur Radioastronomie, Auf dem H\"ugel 69, 53121 Bonn, Germany
\label{bonn}      \and
   Universidade do Porto, Faculdade de Engenharia, Rua Dr.~Roberto Frias, 4200-465 Porto, Portugal
\label{porto}      \and
   School of Physics, University College Dublin, Belfield, Dublin 4, Ireland
\label{dublin}      \and
   Astrophysics Group, Department of Physics \& Astronomy, University of Exeter, Stocker Road, Exeter, EX4 4QL, United Kingdom
\label{exeterAstro}      \and
   Departments of Physics and Astronomy, Le Conte Hall, University of California, Berkeley, CA 94720, USA
\label{ucb}      \and
   European Southern Observatory, Casilla 19001, Santiago 19, Chile
\label{esoc}      \and
   Advanced Concepts Team, European Space Agency, TEC-SF, ESTEC, Keplerlaan 1, NL-2201, AZ Noordwijk, The Netherlands
\label{actesa}      \and
   University of Exeter, Physics Building, Stocker Road, Exeter EX4 4QL, United Kingdom
\label{exeter}      \and
   Fakult\"at f\"ur Physik, Universit\"at Duisburg-Essen, Lotharstraße 1, 47057 Duisburg, Germany
\label{duisburg}      \and
   Instit\"ut f\"ur Astronomie und Astrophysik, Universit\"at T\"ubingen, Auf der Morgenstelle 10, 72076 T\"ubingen, Germany
\label{tuebingen}      \and
   Center for Space and Habitability, Universit\"at Bern, Gesellschaftsstr.~6, 3012 Bern, Switzerland
\label{bern}      \and
   Astronomy Department, University of Michigan, Ann Arbor, MI 48109 USA
\label{umich}      \and
   Institute of Astronomy, University of Cambridge, Madingley Road, Cambridge CB3 0HA, United Kingdom
\label{cam}      \and
   Academia Sinica, Institute of Astronomy and Astrophysics, 11F Astronomy-Mathematics Building, NTU/AS campus, No. 1, Section 4, Roosevelt Rd., Taipei 10617, Taiwan
\label{sinica}      \and
   European Space Agency (ESA), ESA Office, Space Telescope Science Institute, 3700 San Martin Drive, Baltimore, MD 21218, USA
\label{esa}      \and
   Department of Earth \& Planetary Sciences, Johns Hopkins University, Baltimore, MD, USA
\label{jhueps}      \and
   Max-Planck-Institut f\"ur Astrophysik, Karl-Schwarzschild-Stra\ss{}e~1, 85741 Garching, Germany
\label{mpa}      \and
   Excellence Cluster ORIGINS, Boltzmannstra\ss{}e 2, 85748 Garching bei München, Germany
\label{origins}      \and
   Department of Astronomy, Stockholm University, AlbaNova University Center, 10691 Stockholm, Sweden
\label{stockholm}    
}

   \date{Received XXX; accepted YYY}


   \abstract
   {Precise mass constraints are vital for the characterisation of brown dwarfs and exoplanets. Here we present how the combination of data obtained by \textit{Gaia} and GRAVITY can help enlarge the sample of substellar companions with measured dynamical masses. We show how the Non-Single-Star (NSS) two-body orbit catalogue contained in \textit{Gaia} DR3 can be used to inform high-angular-resolution follow-up observations with GRAVITY. Applying the method presented in this work to eight \textit{Gaia} candidate systems, we detect all eight predicted companions, seven of which were previously unknown and five are of a substellar nature. Among the sample is \textit{Gaia} DR3 2728129004119806464 B, which -- detected at an angular separation of \SI{34.01 (15)}{mas} from the host -- is the closest substellar companion ever imaged. In combination with the system's distance and the orbital elements, this translates to a semi-major axis of \SI{0.938 (23)}{AU}. WT 766 B, detected at a greater angular separation, was confirmed to be on an orbit exhibiting an even smaller semi-major axis of \SI{0.676 (8)}{AU}. The GRAVITY data were then used to break the host-companion mass degeneracy inherent to the \textit{Gaia} NSS orbit solutions as well as to constrain the orbital solutions of the respective target systems. Knowledge of the companion masses enabled us to further characterise them in terms of their ages, effective temperatures, and radii via the application of evolutionary models. The inferred ages exhibit a distinct bias towards values younger than what is to be expected based on the literature. The results serve as an independent validation of the orbital solutions published in the NSS two-body orbit catalogue and show that the combination of astrometric survey missions and high-angular-resolution direct imaging holds great promise for efficiently increasing the sample of directly imaged companions in the future, especially in the light of \textit{Gaia}'s upcoming DR4 and the advent of GRAVITY+.
   }


   \maketitle
%

\section{Introduction}

The knowledge of a given substellar companion's mass is key to understanding its formation history, interior evolution, and atmosphere \citep[e.g.][]{Mordasini_luminosity_revisited, Marleau_entropy, DAngelo_giant_planet_formation, Bowler_imaging}. Masses being one of the main exoplanetary observables, they are also of central importance for mapping the demographics of the exoplanet and brown dwarf population (e.g. \citealt{Nielsen_GPI_survey, Vigan_SHINE, Fontanive_HST_survey, Kirkpatrick_IMF, Gratton_JL}). Estimates of a given companion's mass are commonly obtained by applying its approximate age and measured luminosity to evolutionary models (e.g. \citealt{Chabrier_evo_models}; \citealt{Baraffe_Evolutionary_models_2003}; \citealt{Baraffe_Evolutionary_models_2015}). Such indirect mass constraints must, however, be treated with the utmost care since -- in the absence of a large sample of reliably measured masses -- the underlying models are as of yet insufficiently benchmarked and do not always agree with one another \citep{Marley_Lum_of_Young_Jupiters,Mordasini_luminosity_revisited}.
There is thus an urgent need for more directly measured masses of substellar companions.
Examples of such dynamical masses obtained for objects in the brown dwarf mass regime are HD 206893 B \citep{Kammerer_HD206893B}, HD 72946 B \citep{Maire_HD72946B}, and HD 136164 Ab \citep{Balmer_HD136164Ab}. Measured dynamical masses of planetary companion objects, that is, companions with masses below \SI{13}{M_{jup}}, include PDS 70 b \citep{Wang_Constraining_the_nature_of_the_PDS70},
$\beta$ Pictoris c \citep{Nowak_Direct_confirmation}, HR8799 e \citep{Brandt_The_first_dynamical_mass}, AF Lep b \citep{Mesa_AF_Lep_b, Franson_AF_Lep_b, DeRosa_AF_Lep_b}, and HD 206893 c \citep{Hinkley_HD_206893_c}.
A subset of the cited studies make use of the astrometric data collected by ESA's Hipparcos and \textit{Gaia} satellites \citep{Gaia_mission}. Potential proper motion anomalies, the difference in a star's proper motion as measured by the two missions, can be suggestive of the presence of an orbiting companion. Catalogues of such anomalous stars can be found in \citet{Brandt_Hipparcos_Gaia} or \citet{Kervella_stellar_and_substellar}.

Directly imaging substellar companions is still to unfold its great potential in terms of the number of its successful applications (see \citealt{Luvoir_paper, HabEx_paper, Kasper_PCS, Quanz_LIFE}). Yet, even today it can help constrain dynamical masses in a rapid manner. This is the case when it is applied to candidate or confirmed companions whose positions relative to their host stars can be predicted reasonably well and are favourable for such observations.
Such an approach has been made possible by \textit{Gaia} Data Release (DR) 3 \citep{Gaia_DR3}, opening up a new avenue for searches for substellar companions in \textit{Gaia} data that do not include the use of proper motion anomalies. As showcased in this work, the synergies between \textit{Gaia} and GRAVITY \citep{GRAVITY_Collaboration_First_light} makes this instrument ensemble uniquely suited to pursue this new line of attack.

\textit{Gaia} is a space-based survey telescope orbiting the Sun at the Sun-Earth Lagrange point L2, mapping the positions, velocities, spectra, and other characteristics of stars in the solar neighbourhood and beyond. Its astrometric accuracy has been shown to be sufficient to detect the reflex motions that stars exhibit as a consequence of orbiting bodies \citep{Perryman_astrometric_exoplanet_detection, Gaia_Binary_Masses_Table}. In this work, we use the orbits of such host stars to predict the momentary position of the perturbing companion.

The GRAVITY instrument is a near-infrared interferometer at the Very Large Telescope (VLT) on Cerro Paranal. When employed for exoplanet observations, it has been shown to achieve an astrometric accuracy of \SI{50}{\micro as} \citep{Lacour_First_direct_detection}. We used it to follow up on and directly image eight targets identified in the \textit{Gaia} Non-Single-Star (NSS) catalogue.

The paper is structured as follows: How the \textit{Gaia}-informed GRAVITY observations are prepared and conducted is described in Sect.~\ref{section_observations}. The combination of the two data sets as well as further analysis of the GRAVITY observation is presented in Sect.~\ref{section_methods}. The obtained results are then discussed in Sect.~\ref{section_discussion}, and we present our conclusions in Sect.~\ref{section_conclusion}.


\section{Observations} \label{section_observations}

\subsection{Using Gaia as a signpost}
\label{subsection_using_gaia_as_a_signpost}
This work makes use of data obtained by the \textit{Gaia} satellite. The mission's third data release (DR3) \citep{Gaia_DR3} contains the NSS two-body orbit catalogue -- a collection of hundreds of thousands of orbital solutions of point sources of light that can but are not guaranteed to correspond to individual stars. In principle, they can constitute the combination of two or more unresolved sources and are thus referred to as photocentres. A subset of the photocentre orbits are based on astrometric time-series observations taken by \textit{Gaia} over several years. The original astrometric measurements in their raw format are not accessible in DR3. Instead, curated orbital fits are presented as sets of Thiele-Innes (TI) elements, which are directly -- if not linearly -- related to the classical sets of Campbell elements. Additionally, each orbital solution comes with uncertainties associated with the individual elements as well as a correlation matrix. We used \texttt{nsstools}\footnote{https://www.cosmos.esa.int/web/gaia/dr3-nss-tools} \citep{Halbwachs_Gaia_Astrometric_Binary_Star_Processing} to perform the conversions between the TI and Campbell space. 
Dealing with substellar companions, the assumption that the flux seen by \textit{Gaia} originates entirely from the host star is justified for the targets discussed in this work. The validity of such a zero-companion-\textit{Gaia}-flux hypothesis, as we shall henceforth call it, is corroborated by the eventual GRAVITY detection of the selected companions and -- more substantially -- by the later calculation of their synthetic G band magnitudes by means of evolutionary models.
As a consequence of the zero-companion-\textit{Gaia}-flux hypothesis, the photocentre coincides with the position of the host star, and the NSS orbital solution can be interpreted as the star's orbit around its system's centre-of-mass (COM). This enables us to compute the \textit{Gaia}-informed position of the star relative to the COM at any given time. Accordingly, the unresolved companion must be located along the axis connecting the star and the COM and behind the COM as seen from the star. Its exact placement along this axis is determined entirely by the companion-to-host mass ratio. Reflecting on two extreme cases serves as an instructive exercise to make sense of the situation. If the mass ratio approaches \SI{1}{} (that is the companion has the same mass as the host; the initial zero-companion-\textit{Gaia}-flux hypothesis would break down for this extreme case) one would expect the companion to be situated at the opposite side of and at the same distance from the COM as the host. If instead the mass ratio approaches \SI{0}{}, we must expect the companion to be far removed from the host. Thus, lower mass ratios imply greater separations between host and companion.
Both estimates for the host star and companion mass are provided by \textit{Gaia}. From the DR3 Binary Masses Table \citep{Gaia_Binary_Masses_Table}, we retrieve the host star's mass with associated uncertainties. The companion mass is given in the form of an upper and lower limit. Following the reasoning above, by adopting the lower limit as the companion mass we ensure that our eventual host-companion separation is an upper estimate.

\begin{figure*}
        \centering
        \includegraphics[width=0.98\textwidth]{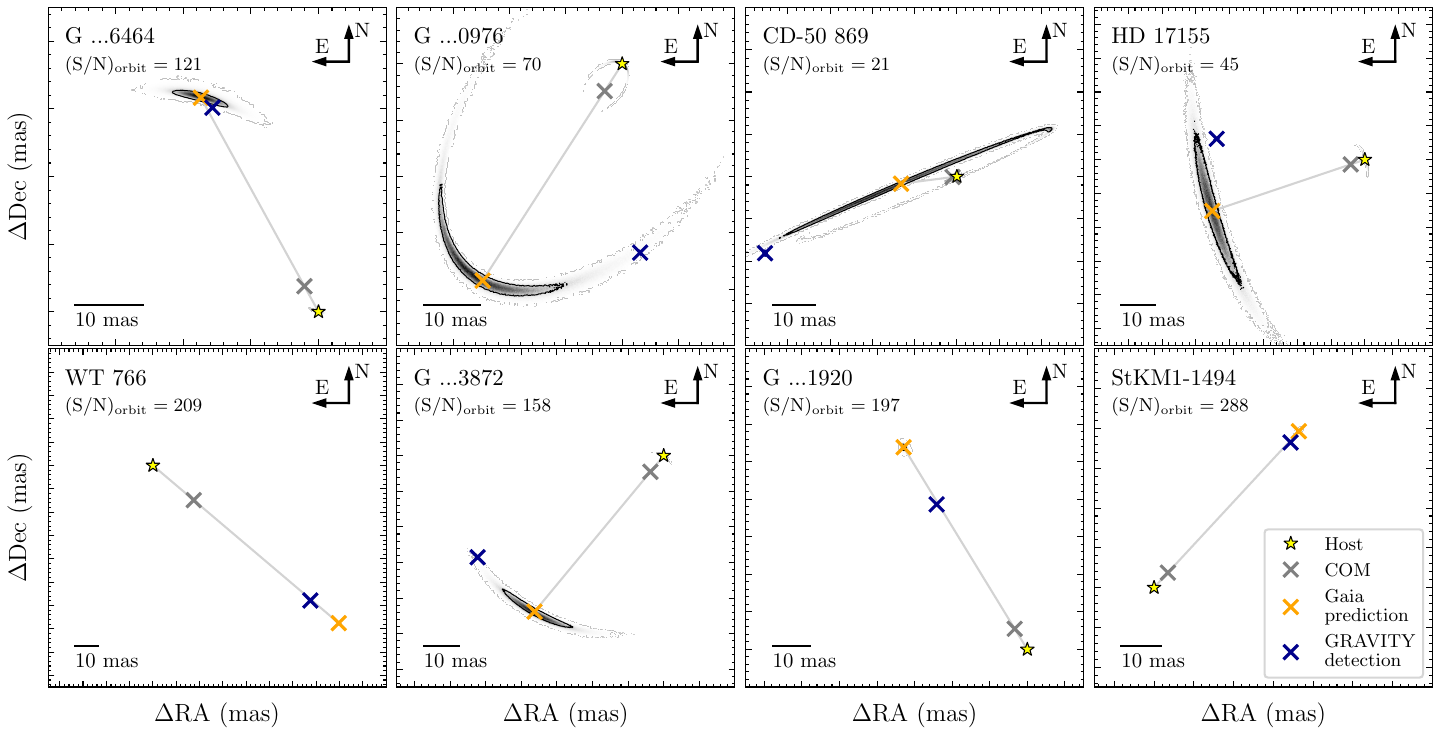}
        \caption{Position probability maps at the time of observation for the observed systems as computed from the respective orbital solutions listed in the \textit{Gaia} NSS catalogue. The orange cross indicates the mean companion position relative to the mean host position, which is marked by the yellow star. The system's COM is shown by the grey cross. Randomising the host and companion positions on the basis of the parameter uncertainties and correlations (while keeping the mass ratio, $q$, fixed to the lower limit; see the main text) yields the predicted spread around the respective mean positions. The contour line encircles \SI{68}{\percent} of the randomised companion positions. The grey axis connects the mean host position with the mean companion position to visualise the separation's dependence on the mass ratio, $q$. The position at which the companion was detected with GRAVITY is shown as the dark blue cross. For targets that were observed more than once, we show the companion position prediction corresponding to the first epoch and the resulting initial detection. The dark blue uncertainty ellipses associated with the GRAVITY detections are too small to be seen at these scales. The ticks along the panel axes are spaced by \SI{10}{mas}. Finally, the significance of the individual orbital solutions, $(\mathrm{S/N})_\mathrm{orbit}$, is given below the target system name in the top-left corner of each panel.}
        \label{figure_ppm_mosaic}
\end{figure*}

Filtering the NSS catalogue according to predefined conditions on different parameters, we can identify favourable targets for a GRAVITY follow-up. Possible filtering conditions are constraints on the RA and Dec positions of the host, its distance, its mass, the companion's lower mass estimate, or the host-companion separation. Age is another possible criterion. It was not applied in the target selection performed for this work, however. We discuss the consequences of this in Sect.~\ref{section_discussion_on_the_G_G_synery}.
The above method enabled us to compile a list of eight target systems viable for a follow up: \textit{Gaia} DR3 2728129004119806464, \textit{Gaia} DR3 4986031970629390976, CD-50 869, HD 17155, WT 766, \textit{Gaia} DR3 4739421098886383872, \textit{Gaia} DR3 4858390078077441920, and StKM1-1494. Among this sample, only the HD 17155 system is known to host a companion (detected via RV monitoring; see \citealt{Barbato_CORALIE_XIX}). For brevity, we shall henceforth shorten the \textit{Gaia} DR3 IDs to the format G ...wxyz, where wxyz are the last four digits of the ID.
To account for the uncertainties of the orbital elements and the correlations between them, we employed a randomisation procedure for each of the companion position predictions. Drawing $N_\mathrm{rand} = \SI{5e5}{}$ different sets of orbital parameters from the multivariate distribution in TI space, converting them to Campbell sets, and evaluating the resulting companion position at a defined time of observation, we obtain $N_\mathrm{rand}$ different position predictions for each companion candidate. These can be visualised in a two-dimensional histogram of the sky plane as shown for each target system in Fig.~\ref{figure_ppm_mosaic}.
We can further quantify the degree of constraint on the companion position prediction by defining a significance metric for the orbital solution. Such a metric can be written as
\begin{align}
    (\mathrm{S/N})_\mathrm{orbit} = \sqrt{\mathbf{TI}_\mathrm{Gaia}^\mathrm{T} \cdot \textbf{COV}_\mathrm{TI,\,Gaia}^{-1} \cdot \mathbf{TI}_\mathrm{Gaia}},
\end{align}
where $\mathbf{TI}_\mathrm{Gaia}$ is the vector containing the TI elements $A$, $B$, $F,$ and $G$ as reported by \textit{Gaia,} and $\textbf{COV}_\mathrm{TI,\,Gaia}$ is the covariance matrix associated with the four elements. The obtained results are presented in the respective panels of Fig.~\ref{figure_ppm_mosaic}.

\subsection{Following-up with GRAVITY}
The targets selected from the \textit{Gaia} NSS two-body orbit catalogue were followed up with GRAVITY between July 2022 and October 2023. Initial observations were conducted on technical time, to demonstrate the validity of the technique. Subsequent observations were conducted within the framework of the ExoGRAVITY Large Programme \citep[ESO ID 1104.C-0651][]{Lacour_ExoGRAVITY_LP} and within a dedicated open time programme (ESO ID 111.253E, PI A.-L. Maire). An observation log containing observation dates, conditions, exposure times and the fibre placement for each target is presented in Table~\ref{table_obs_log}.

All targets were observed using the dual-field on-axis observation mode comprehensively described in \citep{Lacour_First_direct_detection}. To summarise, this technique places the first instrument fibre on the host star to track fringes \citep{Lacour_GRAVITY_Fringe_Tracker}. The science fibre, on the other hand, alternates between the host and the companion.
As this project progressed the observing technique changed slightly.
To begin with, the science fibre was placed at the position at which the companion was predicted to be located (as described in Sect.~\ref{subsection_using_gaia_as_a_signpost}). Later, we decided to position it slightly off-centre to gain in contrast capabilities \citep[this so-called fibre off-pointing technique is comprehensively discussed in][]{Pourre_high_contrast}. The flip side of such an approach is the loss of a fraction of the companion flux. How this throughput loss is corrected for is described in Appendix~\ref{app_section_correcting_for_tp_losses}.

The data reduction was performed using the ESO GRAVITY Instrument Pipeline 1.6.4b1, from which we obtained the astro-reduced byproduct. The astrometry from these files was then obtained using the standard exoplanet dual field data processing described in Appendix A of \citet{2020A&A...633A.110G}. The separations in RA and Dec relative to the hosts at which the companions were detected are shown in Table~\ref{table_astrometry}. Also note that a metrology jump that occurred on Unit Telescope 3 during the observation of the StKM1-1494 system was corrected for.

One of the detections presented in this work (that of G ...6464 B) as well as a preview of the analysis undertaken below was briefly showcased in \citet{Pourre_high_contrast} in order to demonstrate a successful application of the fibre off-pointing observation technique.

\begin{table*}[t]
    \centering
    \tiny
    \caption{Astrometric epochs of the respective companions obtained with GRAVITY.}
    \label{table_astrometry}
    \begin{tabular}{lccccc}
    \toprule
    Target & $t_\mathrm{obs}$ (MJD) & $\Delta$RA (mas) & $\Delta$Dec (mas) & $\rho$ & $(F_c/F_*)_\texttt{2MASS.Ks}$ \\ 
    \midrule
    \midrule
    \object{\textit{Gaia} DR3 2728129004119806464} & 59892.1 & \SI{15.68 (14)}{} & \SI{30.18 (15)}{} & \SI{-0.94}{} & \SI{0.335 (6)}{\percent} \\
    \object{\textit{Gaia} DR3 4986031970629390976} & 59893.1 & \SI{-3.19 (15)}{} & \SI{-33.5 (4)}{} & \SI{0.93}{} & \SI{35.98 (12)}{\percent} \\
    \object{CD-50 869} & 59782.4 & \SI{45.4 (4)}{} & \SI{-18.1 (6)}{} & \SI{0.05}{} & \SI{0.481 (7)}{\percent} \\
    \object{HD 17155} & 59782.3 & \SI{43.8 (7)}{} & \SI{6.2 (4)}{} & \SI{-0.93}{} & \SI{0.472 (5)}{\percent} \\
     & 59810.4 & \SI{44.61 (5)}{} & \SI{2.03 (10)}{} & \SI{-0.68}{} & \SI{0.2890 (15)}{\percent} \\
    \object{WT 766} & 59809.2 & \SI{-67.50 (5)}{} & \SI{-57.89 (5)}{} & \SI{0.48}{} & \SI{6.406 (9)}{\percent} \\
    \object{\textit{Gaia} DR3 4739421098886383872} & 60128.4 & \SI{52.3 (4)}{} & \SI{-28.6 (4)}{} & \SI{0.56}{} & \SI{0.299 (7)}{\percent} \\
     & 60244.3 & \SI{63.28 (5)}{} & \SI{-7.1 (4)}{} & \SI{0.46}{} & \SI{0.173 (19)}{\percent} \\
    \object{\textit{Gaia} DR3 4858390078077441920} & 60156.4 & \SI{24.15 (7)}{} & \SI{38.62 (4)}{} & \SI{-0.44}{} & \SI{1.988 (6)}{\percent} \\
    \object{StKM1-1494} & 60072.2 & \SI{-34.32 (6)}{} & \SI{36.49 (19)}{} & \SI{-0.72}{} & \SI{0.470 (3)}{\percent} \\
    \bottomrule
    \end{tabular}
    \tablefoot{The astrometric companion position components $\Delta$RA and $\Delta$Dec are given relative to the respective host as detected with GRAVITY at the time of observation, $t_\mathrm{obs}$. $\rho$ denotes the correlation between the two astrometric components while the rightmost column shows the companion over host contrast, $(F_c/F_*)$, in the K band after folding with the \texttt{2MASS/2MASS.Ks} filter profile. Note the high K band contrast in the G ...0976 system.}
\end{table*}


\section{Methods}
\label{section_methods}

\subsection{Making Gaia and GRAVITY `talk'} \label{subsection_making_gaia_and_gravity_talk}
The synergies between \textit{Gaia} and GRAVITY are not confined to using the NSS catalogue as a signpost to inform follow-up observations and facilitate detections. By combining the two data sets we can obtain tighter constraints on the orbital elements, break the degeneracy between the host and companion mass that is an inherent feature of \textit{Gaia} NSS orbital solutions and confirm the initial zero-companion-\textit{Gaia}-flux hypothesis.
To this end, we employed a Bayesian approach. In this sense we can interpret the orbital solution provided by \textit{Gaia} as the likelihood information from which the posterior distributions of the orbital elements follow. The GRAVITY detection and concomitant relative astrometry measurement can be viewed as an additional likelihood, upon inclusion of which we can update the posteriors. This amounts to re-sampling the posterior distributions by performing a Markov Chain Monte Carlo (MCMC) run through the multivariate parameter space, spanned by nine free parameters, namely the TI elements $A$, $B$, $F$, and $G$, the eccentricity, $e$, the period, $P$, the time of periastron passage relative to the \textit{Gaia} epoch, $t_\mathrm{p, rel}$, the parallax, $\pi$, and the total mass of the system, $M_\mathrm{tot}$.
To avoid external and non-physical constraints on the individual parameters, we defined a broad, uniform prior range for each of them, virtually ranging from minus infinity to infinity.
Next, a set of $N_w = \SI{100}{}$ walkers were initialised. We assigned a unique set of orbital parameter values to each of them. The actual placement was performed by assuming Gaussian distributions around the \textit{Gaia} NSS orbital elements. To adopt a more conservative initialisation approach and loosen the dependence on the \textit{Gaia} values, we inflated the distributions' standard deviations by a factor of two. This introduces a greater spread ensuring that the prior range is more thoroughly sampled by the initial set of walkers. It should be noted that the total mass is initialised around the host mass estimate, $M_1^\mathrm{Gaia}$, listed in the \textit{Gaia} DR3 Binary Masses Table. Considering the significantly higher mass of the host compared to the companion, this placement appears reasonable.

The MCMC run and actual building of the posterior sampling was performed using \texttt{emcee} \citep{emcee_paper}. At the heart of this procedure lies an ln-likelihood function that reads
\begin{align}
    \ln(\mathcal{L}) = &-\frac{1}{2} \left[ \chi_\mathrm{Gaia}^2 + \chi_\mathrm{GRAVITY}^2 \right],
    \label{equation_ln_likelihood_function}
\end{align}
where $\chi^2$ denotes different chi-squared contributions. The first term, $\chi_\mathrm{Gaia}^2$, compares the individual sets of free parameters sampled by the walkers to the set listed in the \textit{Gaia} NSS catalogue under consideration of the provided covariance matrix. It is thus defined as
{\small
\begin{align}
    \chi_\mathrm{Gaia}^2 = (\bm{\theta}_\mathrm{Sample}-\bm{\theta}_\mathrm{Gaia})^\mathrm{T} \cdot \textbf{COV}^{-1} \cdot (\bm{\theta}_\mathrm{Sample}-\bm{\theta}_\mathrm{Gaia}).
    \label{equation_gaia_term}
\end{align}
}%
Here, $\bm{\theta}_\mathrm{Sample}$ denotes the parameter set vector sampled by the walker, $\bm{\theta}_\mathrm{Gaia}$ is the \textit{Gaia} parameter set as retrieved from the NSS catalogue and $\textbf{COV}$ represents the covariance matrix that follows from the correlation matrix yet again presented in the NSS catalogue.
The second term, $\chi_\mathrm{GRAVITY}^2$, introduces the GRAVITY astrometry into the ln-likelihood function and is defined as
{\small
\begin{align}
    \chi_\mathrm{GRAVITY}^2 = \sum_{n=1}^{N_\mathrm{G}}
    \begin{pmatrix} \bm{\zeta}^\mathrm{Sample}_n - \bm{\zeta}^\mathrm{G}_n \end{pmatrix}^\mathrm{T} \cdot \textbf{COV}^{-1}_n \cdot
    \begin{pmatrix} \bm{\zeta}^\mathrm{Sample}_n - \bm{\zeta}^\mathrm{G}_n \end{pmatrix},
\end{align}
}%
where $N_\mathrm{G}$ is the number of GRAVITY observations and thus astrometric epochs. The vector $\bm{\zeta}^\mathrm{G}_n$ contains the host-companion separation components in RA and Dec measured during the $n$-th GRAVITY observation. $\bm{\zeta}^\mathrm{Sample}_n$, on the other hand, represents the separation components resulting from the parameter set probed by the walker. Finally, $\textbf{COV}^{-1}_n$ describes the covariance matrix associated with the $n$-th GRAVITY observation.
Running the MCMC procedure through a burn-in phase of $N_\mathrm{burn} = \SI{2000}{}$ iterations and building the chain in the actual sampling phase over $N_\mathrm{iter} = \SI{10000}{}$ iterations yields the posteriors.

A word of caution: due diligence is required when applying MCMC methods to data from the \textit{Gaia} NSS two-body orbit catalogue. The orbital solutions contained in the catalogue are presented as sets of TI elements. Related to the classical Campbell elements by a set of non-linear equations \citep{Binnendijk_properties}, the TI elements are known to exhibit strong degeneracies and even circular correlations for circular and quasi-circular orbits. This can lead MCMC procedures astray. Local linear approximation (LLA) methods offer a safe way of utilising \textit{Gaia} NSS orbital solutions \citep{Babusiaux_Gaia_DR3_catalogue_validation}. Information such as the skewness of a posterior distribution or potential multi-modalities, however, is lost when limiting oneself to these inherently Gaussian procedures. Thus, it is desirable to combine the NSS orbital solutions with MCMC methods for such targets where this is possible without misinterpreting the data. We strongly recommend ascertaining the equivalence of LLA and MCMC methods for every target individually.
To this end, we used BINARYS \citep{Leclerc_BINARYS}, a dedicated LLA implementation specifically designed to handle \textit{Gaia} NSS orbit solutions in combination with relative astrometry measurements, to compute an independent set of posteriors under inclusion of all available data. The thus obtained consistent results suggest that the NSS solutions for the targets at hand are suitable to be treated with MCMC procedures.

Finally, it should be noted that the procedure as outlined above cannot be applied to the G ...0976 system. This is due to G ...0976 B's high K band flux relative to its host (see Table~\ref{table_astrometry}). Even though the system's G band flux ratio can be expected to be lower it does not seem reasonable to uphold the zero-companion-\textit{Gaia}-flux hypothesis. Without it, however, we lose the convenient advantage of assuming the photocentre to coincide with the host. In the absence of a measurement of the system's G band flux ratio, it is impossible to convert the photocentre orbit into the host orbit.


\subsection{Orbital refinements}
\begin{figure*}
        \centering
        \includegraphics[width=0.98\textwidth]{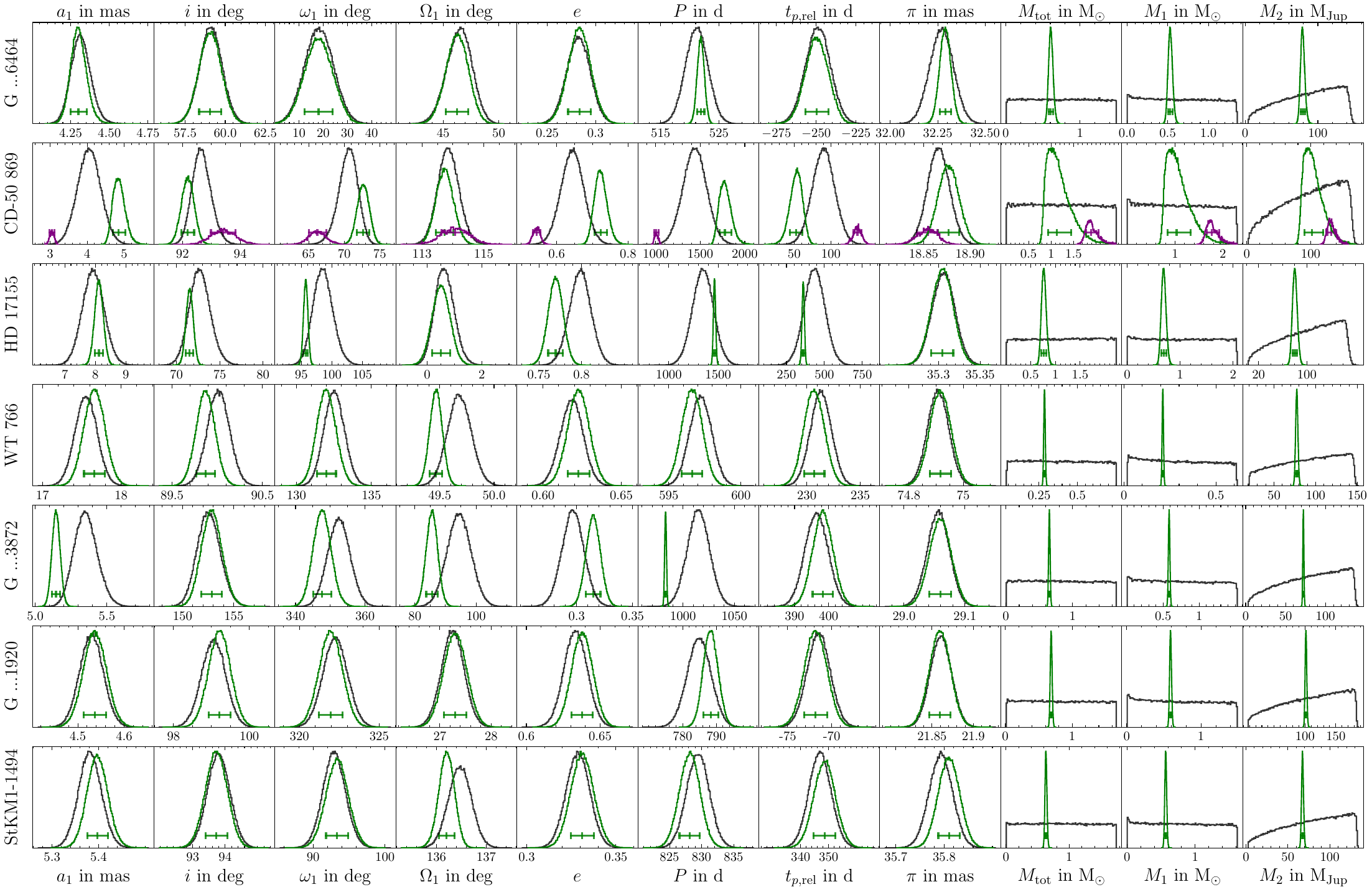}
        \caption{Marginalised posterior distributions of the systems' orbital parameters for the \textit{Gaia}-only run (in black) as well as the \textit{Gaia}-plus-GRAVITY run (in green). Note that the \textit{Gaia}-plus-GRAVITY chain converges on two different orbital solutions for the CD-50 869 system. The bimodal posterior distribution was separated into the preferred (green) and the secondary solution (purple). The indices 1 denote that the shown parameters describe the host star's orbit around the respective system's COM.}
        \label{figure_posteriors_mosaic}
\end{figure*}

\begin{figure*}
        \centering
        \includegraphics[width=0.95\textwidth]{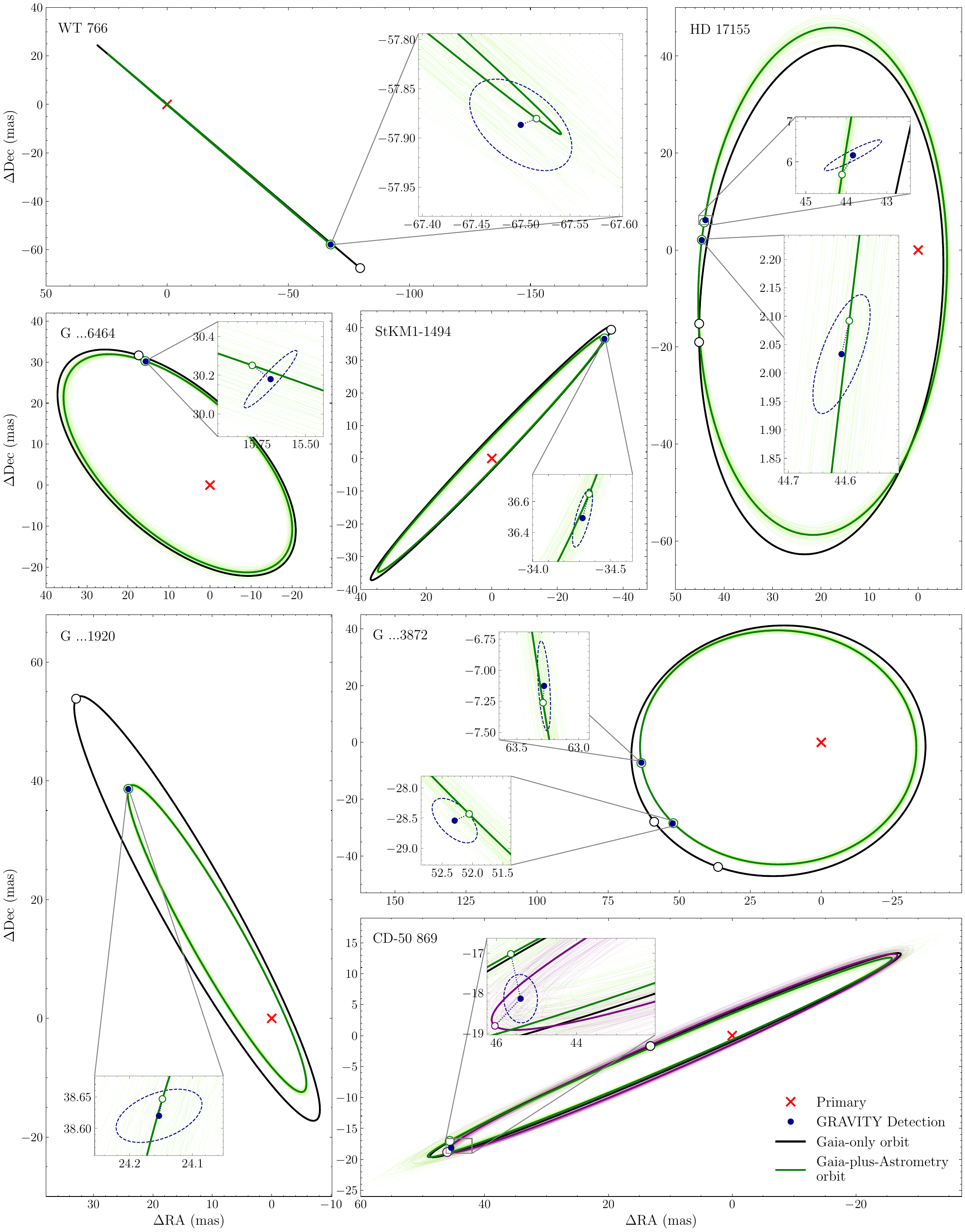}
        \caption{Orbits of the individual companions relative to their respective host. The black line indicates the orbit as suggested by the orbit solution listed in the \textit{Gaia} NSS catalogue (assuming the lower companion mass estimate). The black circle shows where the companion was predicted to be located at the time of observation. The blue dot and its uncertainty ellipse indicate where the GRAVITY detection was made. The thick green orbit is based on the median parameter values of the posterior distributions after combining the \textit{Gaia} and GRAVITY data as described in Sect.~\ref{subsection_making_gaia_and_gravity_talk}. The thin light green lines show \SI{100}{} orbits drawn at random from the posterior distribution. Note that CD-50 869's posterior orbit sampling was separated into a primary and secondary mode, shown here in green and purple, respectively.} 
        \label{figure_orbits_mosaic}
\end{figure*}

The MCMC method outlined in Sect.~\ref{subsection_making_gaia_and_gravity_talk} is applied twice. First, we ran it while only including the \textit{Gaia} term, $\chi_\mathrm{Gaia}^2$, in the ln-likelihood function defined in Eq.~\ref{equation_ln_likelihood_function}. By not considering $\chi_\mathrm{GRAVITY}^2$, we are ignoring the GRAVITY astrometry. We then ran the MCMC procedure taking both terms into account. This let us visualise the constraining power achievable by the inclusion of GRAVITY data, since we can compare the initial \textit{Gaia}-only posteriors to the updated ones. Once the two samplings have been created, we conveyed the individual parameter sets from TI to Campbell space. The resulting full corner plots showing both the initial and the updated posteriors are to be found in Appendix~\ref{app_section_add_tables_and_plots}. The marginalised posterior distributions of the free parameters are shown in Fig.~\ref{figure_posteriors_mosaic}. The inferred median values and associated uncertainties are listed in Table~\ref{table_numerical_results}.
The refinements to the orbital solutions also manifest in the orbits' appearances as projected onto the sky plane. Figure~\ref{figure_orbits_mosaic} shows how the updated orbits compare to the initial \textit{Gaia}-only orbits.

\subsection{Nailing down dynamical masses}
A given host star's movement around its system's COM can conceivably be induced by different combinations of the companion's mass and separation (and when disregarding the zero-companion-\textit{Gaia}-flux hypothesis also by the host-companion flux ratio). For this reason, the \textit{Gaia} astrometric time-series data of the host and the resulting orbital fits by themselves are inherently degenerate in terms of the two bodies' dynamical masses. The columns to the right of Fig.~\ref{figure_posteriors_mosaic} reveal that in addition to enabling significant constraints to be placed on the orbital parameters of target systems, the inclusion of a GRAVITY data point facilitates the breaking of the mass degeneracy. Indeed, one detection suffices to put tight constraints on both absolute masses. Again, the inferred median masses as well as their associated uncertainties are listed in Table~\ref{table_numerical_results}.


\subsection{Photometric characterisation}

Apart from precise relative astrometry measurements between companion and host, the GRAVITY observations of the different target systems also provide so-called contrast spectra in the K band. These measure the flux ratio between companion and host as a function of wavelength. The individual spectra can be found in Fig.~\ref{figure_contrast_spectra_mosaic}. Convolving such a spectrum with a given filter yields the contrast between companion and host as it would appear through the chosen filter. The contrast can then be converted to a magnitude difference. We used \texttt{species} \citep{Stolker_species} to fit the \texttt{BT-Settl (CIFIST)} atmospheric model \citep{Allard_model} to stellar photometry from the literature listed in Table~\ref{table_stellar_mag}. From the resulting fit we then generated a synthetic stellar magnitude in the chosen filter. Finally, the companion magnitude is arrived at by adding the companion-to-host magnitude difference to the synthetic stellar magnitude.

\begin{figure}
        \centering
        \includegraphics[width=0.98\columnwidth]{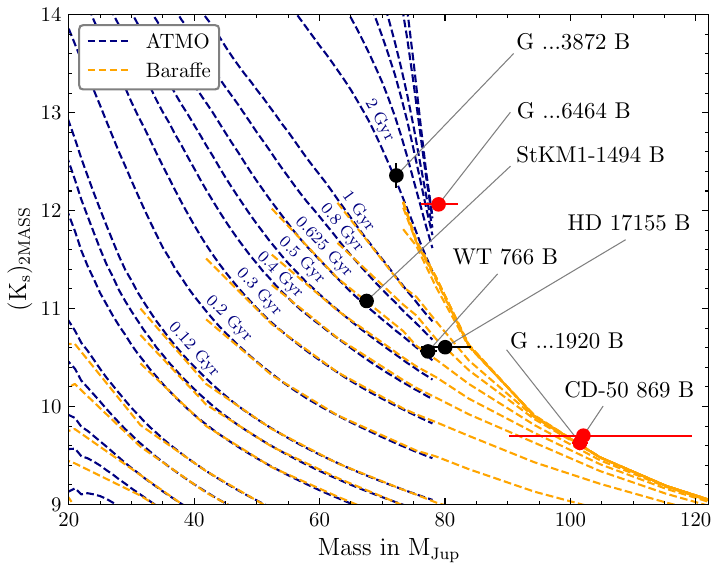}
        \caption{Mass-magnitude plane with the companions detected in this work and isochrones taken from the \texttt{ATMO} and Baraffe model grids in blue and orange, respectively. Both the data points and the isochrones are presented as \texttt{2MASS/2MASS.Ks} filter magnitudes. The three potentially under-luminous companions, G ...6464 B, CD-50 869 B, and G ...1920 B, are marked in red.}
        \label{figure_mass_mag_plane_and_isochrones}
\end{figure}

The knowledge of the companion's dynamical mass alongside its filter magnitude allows us to estimate the object's age by comparing its position on the mass-magnitude plane with isochrones generated using evolutionary models as shown in Fig.~\ref{figure_mass_mag_plane_and_isochrones}.
In other words, this procedure amounts to turning the conventional application of evolutionary models to substellar companions on its head. Typically, they are used to obtain mass estimates from a companion's measured magnitude and its inferred age, which most often is assumed to be similar to that of the stellar association the host belongs to \citep{Bowler_imaging}. As explained above, in our case we can instead obtain constraints on the companion's age from its magnitude and measured mass.
Again, we used \texttt{species} to interpolate isochrones from the \texttt{ATMO} model grid \citep{Phillips_ATMO}. Some companions exhibiting masses above the conventional brown dwarf domain, we included the \citet{Baraffe_Evolutionary_models_2015} model to cover the entire mass range of our target sample.

\begin{figure}
        \centering
        \includegraphics[width=0.98\columnwidth]{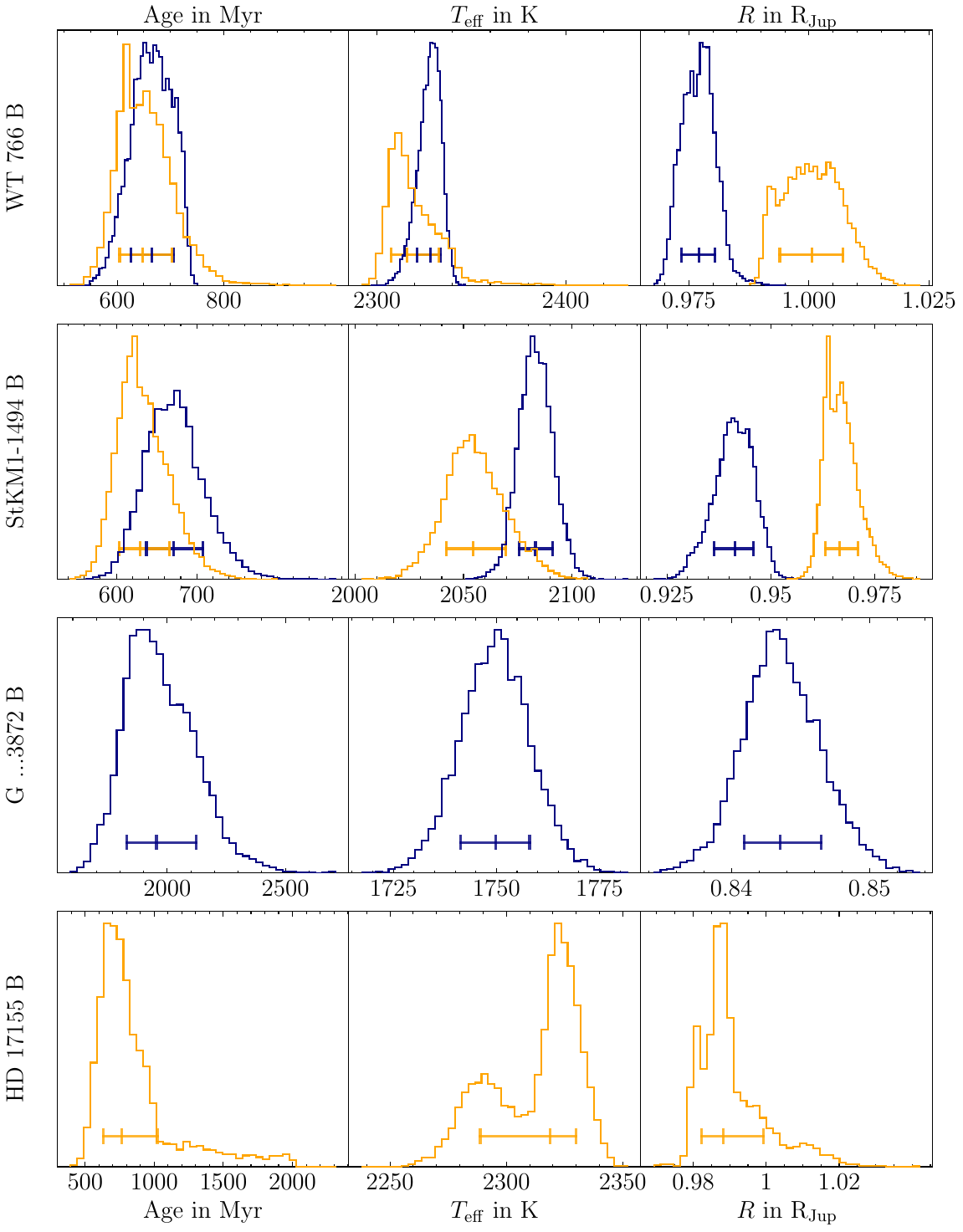}
        \caption{Age, effective temperature, and radius constraints for the individual companions as obtained using the \texttt{ATMO} and Baraffe models in blue and orange, respectively. The horizonal bars show the \SI{1}{\sigma} environment around the median of each distribution.}
        \label{figure_evo_model_parameters_mosaic}
\end{figure}

To eventually arrive at a meaningful age estimates, we propagated the posterior distributions in mass and magnitude by means of a bootstrapping method. To this end we interpolated linearly between the isochrones using \texttt{scipy.interpolate.griddata} \citep{Virtanen_scipy}. This method enables the rapid age determination of mass-magnitude samples randomly drawn from the respective posterior distributions. Computing the ages resulting from $N_\mathrm{boot} = \SI{10000}{}$ mass-magnitude samples for each target yields the age posteriors shown in the left column of Fig.~\ref{figure_evo_model_parameters_mosaic} and listed in Table~\ref{table_numerical_results}. To showcase the power of possessing a high-precision dynamical mass estimate, we can bootstrap the mass and age sampling further into other parameter spaces. Thus, we can repeat the above procedure to obtain posteriors of the companions' effective temperatures and radii. The results are shown in the middle and right column of Fig.~\ref{figure_evo_model_parameters_mosaic} and are listed in Table~\ref{table_numerical_results}.

\begin{figure}
        \centering
        \includegraphics[width=0.98\columnwidth]{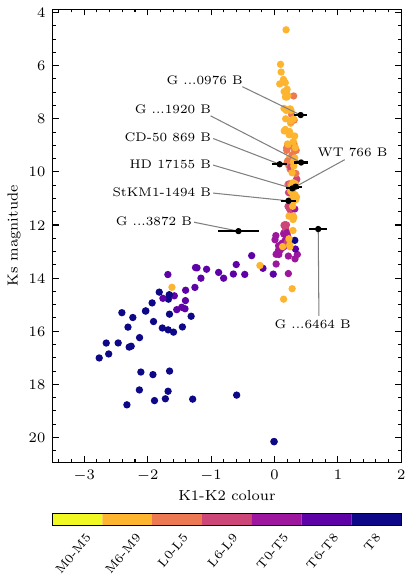}
        \caption{Colour-magnitude diagram showing a template dwarf population compiled in \citet[for details, see their Appendix C]{Bonnefoy_GJ504} as well as the companion sample discussed in this work. The colour indicates the spectral type of a given object.}
        \label{figure_Ks_vs_K2_K1}
\end{figure}

To further contextualise the detected companions we placed them on a colour-magnitude diagram spanned by the \texttt{Paranal/SPHERE.IRDIS\_B\_Ks} filter magnitude and the colour between \texttt{Paranal/SPHERE.IRDIS\_D\_K12\_1} and \texttt{Paranal/SPHERE.IRDIS\_D\_K12\_1}. Such a colour-magnitude diagram, populated with a template dwarfs from the literature (for details see Appendix C of \citealt{Bonnefoy_GJ504}) and the companion sample presented here, can be found in Fig.~\ref{figure_Ks_vs_K2_K1}.

\section{Discussion}
\label{section_discussion}

\subsection{On the Gaia--GRAVITY synergy}
\label{section_discussion_on_the_G_G_synery}
In the previous sections we showed that the orbit solutions contained in the \textit{Gaia} NSS catalogue can be used to inform follow-up observations with a high-precision direct imaging instrument such as GRAVITY. The \textit{Gaia}-based position predictions proved to be sufficiently accurate to facilitate the placement of the GRAVITY science fibre in such a way that the exceedingly small field of view contained the sought-after companion. In this way, we were able to detect and confirm all eight predicted companions.

The complete lack of non-detections is worthy of a short discussion. As the reader will recall, young age was not used as a criterion for target selection. Thus, one would naïvely assume that such an allowance for old targets would cause some of the observations to result in non-detections of the respective companions since the companion-to-host contrast decreases with age. A simple estimation of the companion magnitude at high ages resolves the conundrum. Given the inferred masses, we could compute the magnitudes from evolutionary models. Furthermore, we could compute the limiting magnitude up to which detection with GRAVITY can be achieved for each target system. To this end, we evaluated the GRAVITY contrast curve \citep{Pourre_high_contrast} at the detection separation of each companion and converted the resulting limiting contrasts to a magnitude difference. In combination with the respective K band magnitudes of the hosts this yields the limiting companion magnitudes. Comparison with the companion magnitudes from the evolutionary models reveals that they are all observable independent of their ages. This is a consequence of investigating companions located towards the upper end of the brown dwarf mass range. We are dealing with objects that are sufficiently bright to facilitate successful detection regardless of their age. This is a luxury not afforded to those who plan to apply the procedures outlined here to less massive and therefore fainter companion targets. In such cases one is forced to include age as a selection criterion.

Substellar companions in the NSS catalogue typically reside at separations of a few dozen \SI{}{mas}. This makes follow-up observations with classical direct imaging instruments such as SPHERE \citep{Beuzit_SPHERE} or GPI \citep{Macintosh_GPI} challenging for the vast majority of these objects due to the inner working angle of their coronagraphs. It is only by the means of GRAVITY and its unique angular resolution that these low-separation companions can be accessed. This will be even more relevant for the numerous \textit{Gaia}-inferred companions in the planetary-mass regime, which are predicted to be found at even smaller separations \citep{Perryman_astrometric_exoplanet_detection}.
To illustrate this point, it is worth noting that among the sample of detected companions presented here is \textit{Gaia} DR3 2728129004119806464 B, which, observed at a radial separation of \SI{34}{mas} from the host star, is the closest ever directly detected substellar companion. The inferred companion orbit puts the semi-major axis at \SI{0.938 (23)}{AU}.
Another noteworthy example is that of WT 766 B's orbit, which exhibits a semi-major axis of \SI{0.676 (8)}{AU}. The semi-major axes of the entire companion sample are presented in Table~\ref{table_sma_i_chi_squared}. 

The independent companion detections by GRAVITY provide a rare opportunity to validate and assess the orbital solutions published in the \textit{Gaia} DR3 NSS two-body orbit catalogue. One way of doing so is by assessing the statistic validity of the \textit{Gaia} uncertainties. To this end we compared the companion position prediction based on the NSS orbital solutions to the GRAVITY detection. Since the radial separation component depends on the initially unconstrained mass ratio between companion and host, we removed it from the analysis by only considering the position angle component, $\varphi$. If the \textit{Gaia} uncertainties are statistically robust approximately $\SI{68}{\percent}$ of the detections should be made at position angles within a $\SI{1}{\sigma}$ confidence interval around the median of the \textit{Gaia}-based position angle prediction, $\varphi_\mathrm{med}$. Figure~\ref{figure_inflation_factors} shows said position angle distribution for each target system as well as the respective GRAVITY detection position angle, $\varphi_\mathrm{det}$. By taking the absolute difference, $\Delta \varphi = |\varphi_\mathrm{det} - \varphi_\mathrm{med}|$, between the distribution's median and the detection value and scaling it by the respective uncertainties, $\sigma_{\varphi,\,\mathrm{det}}$ and $\sigma_{\varphi,\,\mathrm{med}}$, according to
\begin{align}
    I = \frac{\Delta \varphi}{\sqrt{\sigma_{\varphi,\,\mathrm{det}}^2 + \sigma_{\varphi,\,\mathrm{med}}^2}},
\end{align}
we can compute an inflation factor, $I$, that quantifies by how much the \textit{Gaia} uncertainties need to be `blown up' to make the $\SI{1}{\sigma}$ position angle confidence interval encompass the detection.
The resulting inflation factors for the individual systems are showcased in Table~\ref{table_sma_i_chi_squared}. They indicate that -- with the exception of G ...6464 B -- all companions were detected outside the respective \textit{Gaia}-based $\SI{1}{\sigma}$ position angle confidence interval, suggesting that the \textit{Gaia} uncertainties are underestimated. We note that as can be seen in Fig.~\ref{figure_inflation_factors} the inflation factor found for the CD-50 869 system results from a non-Gaussian position angle distribution rendering the value unusable.

\subsection{Orbital refinements and dynamical masses}
By means of a Bayesian inference framework conducted using an MCMC procedure, we were able to combine the orbit solution provided by \textit{Gaia} with the GRAVITY astrometry measurement. Visual inspection confirmed that, apart from the mass chains, all chains of the \textit{Gaia}-only runs as well as all chains of the \textit{Gaia}-plus-GRAVITY runs converged for every target system. The inclusion of the GRAVITY detection resulted in tighter constraints on the orbital elements and the breaking of the mass degeneracy inherent to the NSS orbit solution. With the exception of CD-50 869 B and G ...1920 B, which have both been shown to be of stellar nature, the resulting dynamical masses lie within the respective mass constraints listed in the DR3 Binary Masses Table \citep{Gaia_Binary_Masses_Table}. While the dynamical mass of CD-50 869 B carries a relative uncertainty of \SI{15}{\percent}, those of the other targets vary between \SI{0.9}{} and \SI{5}{\percent}.

Furthermore, the posterior sampling of the CD-50 869 system is of bimodal nature. A possible explanation of this circumstance can be found in the geometry of the orbit and epoch of the GRAVITY detection. In order to minimise leakage of the host star flux into the science fibre and the resulting speckle noise it was decided to observe this companion at its furthest separation from the host. In combination with the orbit being almost edge-on, the MCMC chain converges on two different solutions that place the companion before and after the projected turning point, respectively. Such a configuration serves to show the limitation of this method. By observing at the largest separation and thereby maximising the chances of detection, one runs the risk of obtaining an inconclusive posterior due to bi- or even multi-modality. It should be noted, however, that the acquisition of an additional GRAVITY epoch of the companion at a different position along its orbit would likely kill the bimodality and pin down the orbital solution at or close to the primary mode obtained here.
Finally, when allowing for additional evidence from proper motion anomaly studies, CD-50 869's entry in the catalogue put forward in \citet{Kervella_stellar_and_substellar} suggests a host star mass of \SI{0.90 (4)}{M_\odot}. This estimate is covered by the primary mode's host mass constraint of $1.03^{+0.32}_{-0.19}$ \SI{}{M_\odot}. Moreover, using surface brightness to colour relations and the system's parallax they also obtain a host star radius estimate of \SI{0.87 (4)}{R_{Sun}}, which is incompatible with the secondary mode's large host mass. 
The only other system presented here that is also contained in the \citet{Kervella_stellar_and_substellar} proper motion anomaly catalogue is HD 17155.

While more constrained, on the whole the updated orbit posteriors visualised in Fig.~\ref{figure_posteriors_mosaic} agree with the \textit{Gaia} NSS orbital solutions. The largest discrepancy between the posteriors before and after inclusion of the GRAVITY astrometry can be seen in the orbital period of G ...3872 B. We note that this period value is particularly large, suggesting that the period estimate presented in the NSS catalogue might have been underestimated due to insufficient orbital coverage.

\begin{figure}
        \centering
        \includegraphics[width=0.9\columnwidth]{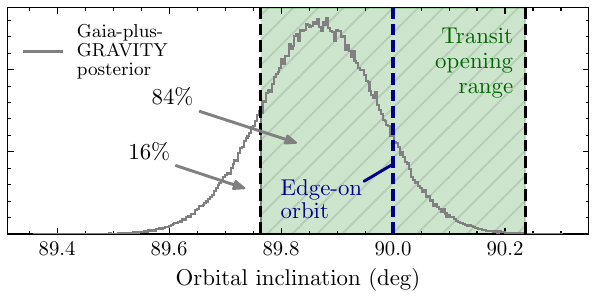}
        \caption{WT 766 B's \textit{Gaia}-plus-GRAVITY inclination posterior compared to the angular transit opening range indicated by the hatched green region. The dashed blue line shows a perfectly edge-on orbit. The percentage values specify the posterior sampling fractions falling within and outside the opening range.}
        \label{figure_transit_prob}
\end{figure}

The almost exactly edge-on orbit of WT 766 B was already indicated by the orbital solution listed in the \textit{Gaia} NSS catalogue. The updated orbit elements substantiate this orbit configuration and suggest the possibility of observing this substellar companion in transit.
The transit probability can be calculated on the basis of the maximum opening angle, $\vartheta$. If a given companion's orbit exhibits an inclination within the range $\SI{90}{^\circ}-\vartheta < i < \SI{90}{^\circ}+\vartheta$ it will show a transit. Based on the derivation in \citet{beatty_seager_transit_probabilities} the maximum opening angle, $\vartheta$, for an eccentric companion orbit can be computed via
\begin{align}
    \vartheta = \arcsin \left(\frac{R_*}{a} \cdot \frac{1 + e\sin(\omega)}{1-e^2} \right),
\end{align}
where $R_*$ is the stellar radius and $a$, $e$, and $\omega$ are the companion orbit's semi-major axis, eccentricity, and argument of periastron, respectively. For WT 766 B this yields a maximum opening angle of $\vartheta = \SI{0.236 (7)}{^\circ}$. How this compares to the companion's inclination posterior is visualised in Fig.~\ref{figure_transit_prob}. Integrating the posterior within the angular transit opening range around \SI{90}{^\circ} yields a transit probability of \SI{84}{\percent}.

According to the orbital solution posterior the next transit will occur in February 2025. The symmetry of the orbit as projected onto the sky plane makes it impossible to determine whether this will constitute a primary or secondary eclipse, however. Upon observation it should be fairly obvious which phenomenon is observed, breaking the degeneracy in the orbit symmetry. Alternatively, an additional RV measurement of the system on top of the one listed in \textit{Gaia} DR3 should enable the derivation of the systematic velocity and in turn resolve the orbit's directional degeneracy. In any case, such an observation would constitute the first application of imaging and transit techniques to the same object.

In the respective panels of Fig.~\ref{figure_orbits_mosaic} we can observe that the orbital constraints resulting from the inference procedure also manifest in changes to the geometry of some of the orbits.
To prepare for computing reduced chi-squared values of the refined orbits we ran an additional gradient descent procedure starting at the most probable parameter sample as determined by the MCMC run. The minima identified by the gradient descent procedure were then used to compute the chi-squared values presented in Table~\ref{table_sma_i_chi_squared}.

\begin{table}[H]
    \centering
    \caption{Semi-major axes, inflation factors and reduced chi-squared values for the detected companions.}
    \label{table_sma_i_chi_squared}
    \resizebox{\columnwidth}{!}{%
    \begin{tabular}{lccc}
    \toprule
    Target & $a$ in \SI{}{AU} & $I$ & $\chi_\mathrm{red}^2$ \\ 
    \midrule
    \midrule
    \textit{Gaia} DR3 2728129004119806464 B & \SI{0.938 (23)}{} & 0.4 & 0.04 \\
    \textit{Gaia} DR3 4986031970629390976 B & --\tablefootmark{a} & 2.9 & --\tablefootmark{a} \\
    CD-50 869 B & \SI{2.73 (28)}{} & 0.2\tablefootmark{b} & 3.67 \\
    HD 17155 B & \SI{2.07 (7)}{} & 1.6 & 2.52 \\
    WT 766 & \SI{0.676 (8)}{} & 2.9 & 1.84 \\
    \textit{Gaia} DR3 4739421098886383872 B & \SI{1.500 (19)}{} & 3.0 & 2.19 \\
    \textit{Gaia} DR3 4858390078077441920 B & \SI{1.265 (16)}{} & 1.4 & 0.54 \\
    StKM1-1494 B & \SI{1.329 (14)}{} & 1.3 & 0.92 \\
    \bottomrule
    \end{tabular}
    }
    \tablefoot{
    The semi-major axes, $a$, shown here result from the \textit{Gaia}-plus-GRAVITY orbital posteriors. The inflation factors, $I$, are required to make the \SI{1}{\sigma} position angle confidence interval encompass the detection. The presented reduced chi-squared values of the best-fit orbital solution were found by means of a gradient descent from the most probable parameter set in the MCMC posterior sampling.
    \tablefoottext{a}{Note that G ...0976 B lacks values for the semi-major axis as well as the reduced chi-squared since the posterior could not be sampled for this companion.}
    \tablefoottext{b}{The inflation factor found for the CD-50 869 system results from a non-Gaussian position angle distribution rendering the value unusable.}
    }
\end{table}

\subsection{Inferring ages from evolutionary models}
\label{section_inferring_ages_from_evo}

The first notable feature of Fig.~\ref{figure_mass_mag_plane_and_isochrones} is the potential under-luminosity observed in three of the detected companions, namely G ...6464 B, CD-50 869 B, and G ...1920 B. For two of these (G ...6464 B and CD-50 869 B) the error bars on their dynamical mass estimates still overlap with the oldest isochrones. This is not the case for G ...1920 B, which is significantly removed from the isochrones. Under-luminosity in old brown dwarf companions is a known problem in the literature (e.g. \citealt{Brandt_improved}). Invoking the possibility of companion binarity offers a potential explanation as to the observed lack in luminosity. A companion to the companion would distort the mass of the primary companion since the treatment as put forward here accounts for the putative pair of bodies as a single point mass located at the astrometric position measured by GRAVITY. This would move the respective companion's data point in Fig.~\ref{figure_mass_mag_plane_and_isochrones} towards the right into a domain where it appears to be under-luminous when instead its mass was misjudged by assuming the companion system to be a single body.

The presence of a secondary companion could be observable by obtaining several GRAVITY epochs. If the astrometric data can be shown to deviate from the expected orbit and to follow a trajectory that exhibits features reminiscent of epicycles, this would strongly suggest companion binarity. The magnitude of such a feature would depend on the mass ratio between the primary and the secondary companion, which can be estimated to the first order by measuring the excess mass in Fig.~\ref{figure_mass_mag_plane_and_isochrones}. Such a follow-up study could not be conducted within the framework of this work since the host-companion separations of the targets at hand have already dropped into a range at which even GRAVITY observations are impossible. Once the separations have sufficiently increased again, a dedicated follow-up should be able to shed light on the question of companion binarity.

A note on the filter magnitudes shown in Fig.~\ref{figure_mass_mag_plane_and_isochrones}: to present the isochrones of both evolutionary models in the same filter system, we converted the \texttt{ATMO} isochrones from \texttt{APO/NSFCam.MKO\_K} to \texttt{2MASS/2MASS.Ks} magnitudes (for details on the conversions between the two filter systems, see \citealt{Hawarden_UKIRT} and \citealt{Carpenter_2MASS}). The slight divergence of the isochrones towards lower masses can be explained by the fact that the filter system conversion was derived using stellar photometry. It should be noted, however, that the conversion was only performed for visualisation purposes within Fig.~\ref{figure_mass_mag_plane_and_isochrones}.
The subsequent bootstrapping, which was conducted in the \texttt{APO/NSFCam.MKO\_K} and \texttt{2MASS/2MASS.Ks} filter magnitude space for the \texttt{ATMO} and Baraffe models, respectively, is therefore not affected by any potential isochrone divergence.

Since WT 766 B and StKM1-1494 B lie in a the region of the mass-magnitude plane where both evolutionary models are defined, we were able to use both to infer age, effective temperature and radius.
Figure~\ref{figure_evo_model_parameters_mosaic} and Table~\ref{table_numerical_results} show that the ages inferred using both models match. 
While the effective temperature posteriors agree for WT 766 B this is not the case for StKM1-1494 B. Furthermore, the radius posteriors from both models disagree for both targets. For G ...3872 B and HD 17155 B only one model could be used for the characterisation. Due to their placement on the mass-magnitude plane, no such analysis could be conducted for the three potentially under-luminous targets G ...6464 B, CD-50 869 B, and G ...1920 B.

The clustered and thus degenerate nature of isochrones describing older objects within the mass-magnitude plane is reflected in the age posteriors of some of the targets. Their samplings are skewed towards older ages.
In general the obtained ages for our target sample companions are remarkably low. This comes as a surprise seeing that the stellar age distribution in the solar neighbourhood peaks at a significantly higher value (at $\sim$\SI{5}{Gyr} according to \citealt{gondoin_age}). This bias towards young ages could be caused by an observational overestimation of the companion flux amounting to a photometric calibration error of the GRAVITY data.
Investigation into whether or not this is the cause of the observed bias is outside the scope of this work. If confirmed, however, correcting for the resulting magnitude offset would exacerbate the aforementioned under-luminosity observed in three of the companions.

Having acquired the companions' dynamical masses as well as their ages, we could apply the same evolutionary models as above to generate synthetic companion magnitudes in different wavelength bands. This serves two goals. On the one hand it enables us to verify that our initial zero-companion-\textit{Gaia}-flux hypothesis was justified, closing the loop and ascertaining that the mass constraints on the hosts and companions are trustworthy. On the other hand these magnitudes will be informative for potential future follow up observations of the companions with other instruments working in different wavelength ranges. The respective synthetic companion magnitudes are listed in Table~\ref{table_synth_comp_mags}. Inspection of the G band flux ratio of the target systems for which it could be computed affirms that they are indeed negligible. 

Figure~\ref{figure_Ks_vs_K2_K1} establishes that the majority of new companions presented here are spread out along the L-dwarf branch with G ...0976 B, G ...1920 B, and CD-50 869 B consistent with M-L transition bodies \citep{Kirkpatrick_M_L_type}. G ...3872 B and G ...6464 B are the only companions discussed here that appear removed from the literature population. In the case of G ...6464 B this could constitute another hint at companion binarity. Alternatively, as \citet{Faherty_signatures} point out, the strong red colour of G ...6464 B might indicate a thick cloud deck in its atmosphere.
G ...3872 B on the other hand sits above the L-T \citep{Kirkpatrick_L_T_type} transition possibly indicating a methane rich atmosphere.

Finally, it is worth remembering that we detected an additional companion, which -- judging by its large K band flux relative to its host -- is likely of stellar nature. The high contrast was the reason G ...0976 B could not be included in the analysis since the crucial zero-companion-\textit{Gaia}-flux hypothesis could not be upheld. Since the orbit listed in the \textit{Gaia} NSS two-body orbit catalogue did facilitate a successful detection, however, there is every reason to expect that such a \textit{Gaia}-informed observation can be repeated. Obtaining additional epochs of this companion will enable further constraints to be placed on its orbit and thereby its mass, potentially solving the question as to why it is so bright in the K band.


\section{Conclusion}
\label{section_conclusion}

In this work we have demonstrated the synergies between \textit{Gaia} and GRAVITY by showcasing how the astrometry-based orbital solutions contained in the NSS two-body orbit catalogue can be used to inform high-angular-resolution follow-up observations. We detected and confirmed eight companions, of which seven were previously unknown. Every conducted observation resulted in the successful detection at or close to the predicted position of the inferred companion.
It is important to note that the unprecedentedly precise nature of the astrometric data collected by \textit{Gaia} makes it difficult to follow up on or challenge using other currently available instruments. In most cases, the high contrasts and small angular separations between hosts and companions as suggested by the orbits listed in the NSS catalogue are inaccessible to today's direct imaging facilities. From this point of view, GRAVITY offers a unique opportunity to test the orbits put forward in the NSS two-body orbit catalogue. We can conclude that the results presented here -- both in terms of the 100\% success rate in detecting the inferred companions and in terms of how the two data sets can be combined -- constitute an independent validation of the \textit{Gaia} NSS orbital solutions as well as the \textit{Gaia} binary mass estimates.
Applying the \textit{Gaia}--GRAVITY ensemble to companion candidates of planetary nature will be an intriguing next step in exploring the two instruments' joint potential.

Combination of the \textit{Gaia} and GRAVITY data by means of a Bayesian approach revealed more refined orbital solutions as well as tight constraints on the host and companion masses. They suggest that five companions of the sample can be categorised as brown dwarfs, while two, CD-50 869 B and G ...1920 B, are of a stellar nature. The same can be expected for G ...0976 B.

The GRAVITY K band spectra were investigated to further characterise the companions. To this end, evolutionary models were used to obtain constraints on their age, effective temperature, and radius. The possibility of probing for companion binarity by means of a high-resolution astrometric time series with GRAVITY is discussed here and should be considered for the potentially under-luminous targets G ...6464 B, CD-50 869 B, and G ...1920 B. Comparison with a template population from the literature further contextualises the companion sample and indicates that they are consistent with L- to T-type brown dwarfs.

The potential of this method in confirming candidates suggested by \textit{Gaia} astrometry cannot be overestimated. Rapid direct detections of the inferred companions render the breaking of the mass degeneracy between host and companion a matter of routine. The resulting tight constraints on the companion masses can shed light on formation processes and interior evolution.

The procedures showcased in this work are not confined to brown dwarf companions but can readily be applied to planetary candidates as well. Such candidates are still scarce in DR3 but should be forthcoming in DR4, which is expected to be published in 2026. The astrometric time-series data to be released therein will amount to a treasure trove of planetary candidates accessible with GRAVITY. And finally, with the imminent arrival of GRAVITY+, the possibility of accessing companions at even smaller contrasts bodes well for the future synergies between \textit{Gaia} and GRAVITY as well as direct imaging in general.


\begin{acknowledgements}
This work is based on observations collected at the European Southern Observatory under ESO programmes 60.A-9102, 1104.C-0651 and 111.253E.   
SL acknowledges the support of the French Agence Nationale de la Recherche (ANR), under grant ANR-21-CE31-0017 (project ExoVLTI).
G-DM acknowledges the support of the DFG priority program SPP 1992 ``Exploring the Diversity of Extrasolar Planets'' (MA~9185/1) amd from the Swiss National Science Foundation under grant 200021\_204847 ``PlanetsInTime''.
Parts of this work have been carried out within the framework of the NCCR PlanetS supported by the Swiss National Science Foundation.
J.J.W., A.C., and S.B.\ acknowledge the support of NASA XRP award 80NSSC23K0280.
This work has made use of data from the European Space Agency (ESA) mission
{\it Gaia} (\url{https://www.cosmos.esa.int/gaia}), processed by the {\it Gaia}
Data Processing and Analysis Consortium (DPAC,
\url{https://www.cosmos.esa.int/web/gaia/dpac/consortium}). Funding for the DPAC
has been provided by national institutions, in particular the institutions
participating in the {\it Gaia} Multilateral Agreement.
This research has made use of the Jean-Marie Mariotti Center \texttt{Aspro}
service \footnote{Available at http://www.jmmc.fr/aspro}.
This publication makes use of data products from the Wide-field Infrared Survey Explorer, which is a joint project of the University of California, Los Angeles, and the Jet Propulsion Laboratory/California Institute of Technology, funded by the National Aeronautics and Space Administration. \\
Finally, we would like to thank Timothy Brandt for his thorough and insightful referee report. His comments and suggestions have substantially improved the quality of this paper.

\end{acknowledgements}


%
%




\bibliographystyle{aa} 
\bibliography{refs.bib} 


\begin{appendix}

\section{Correcting for throughput losses}
\label{app_section_correcting_for_tp_losses}

\begin{figure}[h]
        \centering
        \includegraphics[width=0.98\columnwidth]{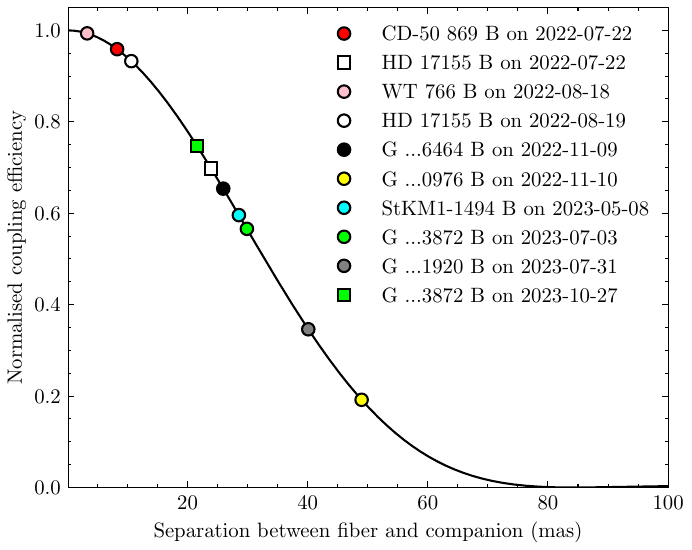}
        \caption{Normalised coupling efficiency, $\gamma$, as a function of the angular separation between the fibre centre and the observed body.}
        \label{figure_tp_losses}
\end{figure}

When observing targets in the dual-field on-axis mode of GRAVITY, situations can arise where the companion body is separated from the centre of the science fibre by a certain angular distance. This can be case when the companion's position relative to the host cannot be constrained well enough to inform the proper placement of the fibre or when the observer deliberately moves the fibre in such a way that the target is not located at its centre anymore. This fibre off-pointing technique \citep{Pourre_high_contrast} can facilitate contrast capability gains rendering it especially useful for such observations where one expects unfavourable contrast conditions between the companion and host. For both these reasons the observations conducted for this work exhibit angular separations between the eventually detected companions and the fibre-centre. Following the procedure outlined in Appendix A of \citet{Wang_Constraining_the_nature_of_the_PDS70}, we computed the normalised coupling efficiency, $\gamma$, as a function of the angular separation from the fibre centre. The relation as well as the respective companions' separations are presented in Fig.~\ref{figure_tp_losses}.
Dividing the observed contrast spectra by the respective coupling efficiency, $\gamma$, we can correct for the ensued throughput losses.

\section{Inflation factors for \textit{Gaia} uncertainties}
\label{app_section_inflation_factors}

\begin{figure}[h]
        \centering
        \includegraphics[width=0.78\columnwidth]{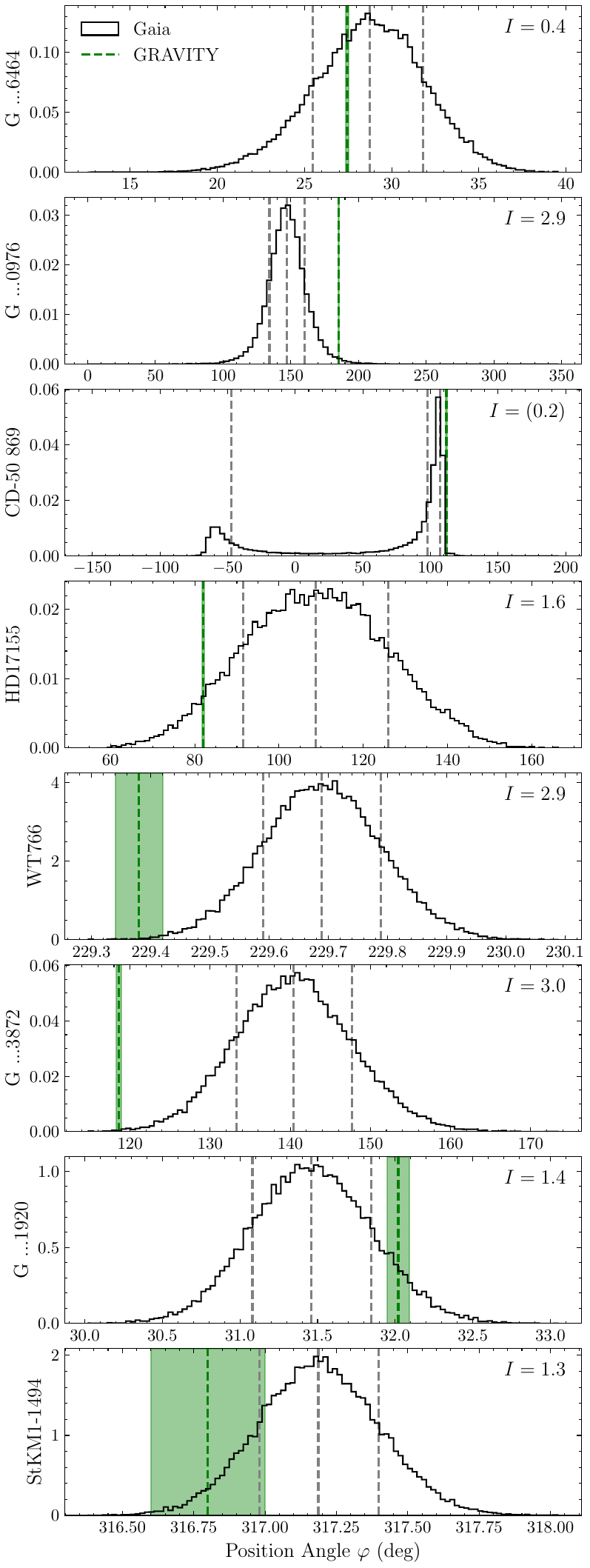}
        \caption{Position angle distributions of the individual companions as predicted from the \textit{Gaia} NSS two-body orbit solutions as well as the position angles at which the respective companions were detected using GRAVITY (in green). The inflation factor, $I$, required to make the \SI{1}{\sigma} confidence interval encompass the detection is given in the top-right corner of each panel. Note that the distribution's non-Gaussian shape in the panel for the CD-50 869 system renders the resulting inflation factor unusable.}
        \label{figure_inflation_factors}
\end{figure}

As discussed in Sect.~\ref{section_discussion_on_the_G_G_synery}, the companion position angle predictions based on the \textit{Gaia} NSS two-body orbit catalogue in combination with the true position angle measured with GRAVITY can be used to assess how robust the \textit{Gaia} uncertainties are. In Fig.~\ref{figure_inflation_factors} we present the comparison between prediction and detection for each target system.

\section{Additional tables and plots}
\label{app_section_add_tables_and_plots}
\begin{figure*}[h]
        \centering
        \includegraphics[width=0.98\textwidth]{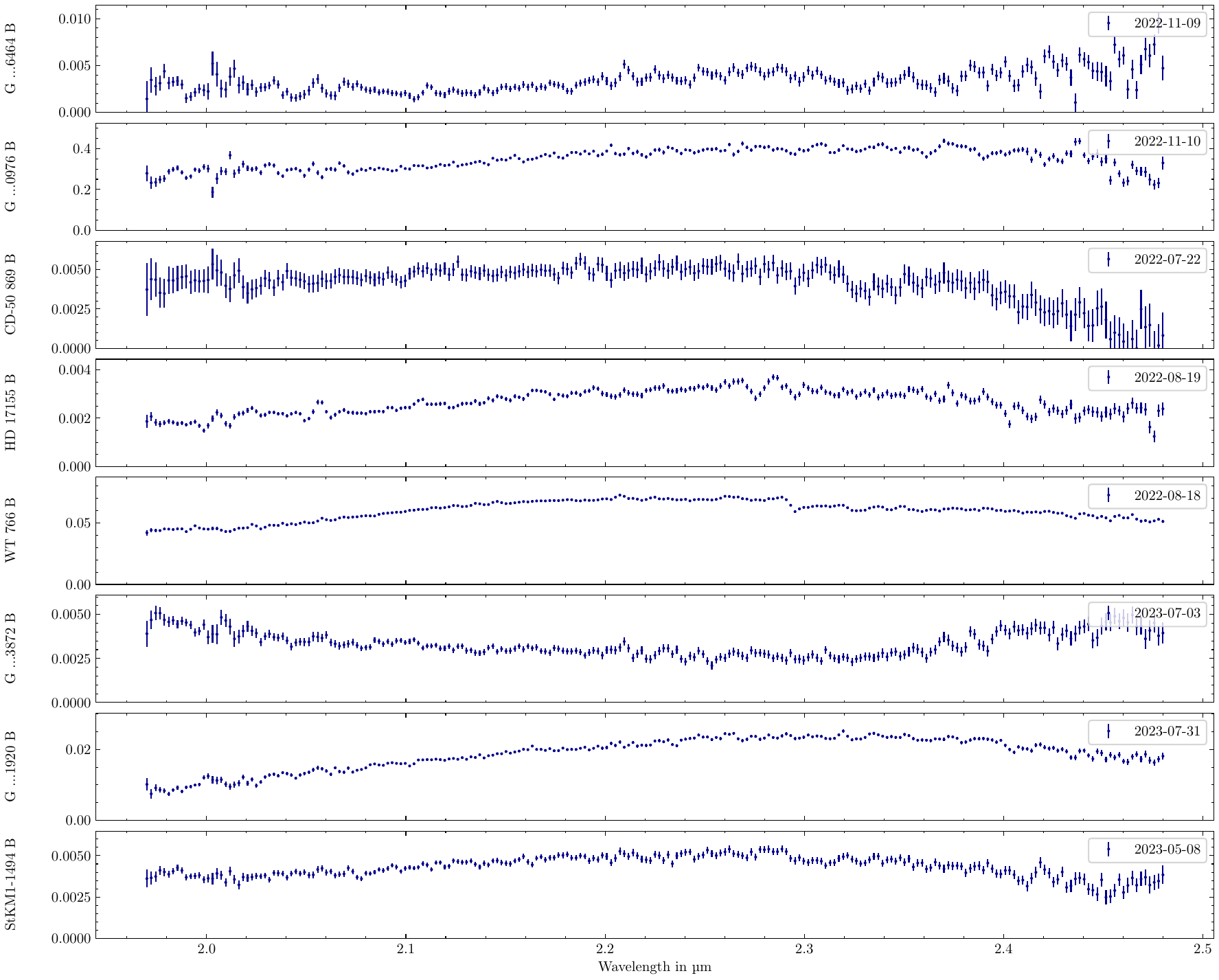}
        \caption{GRAVITY contrast spectra between the companion and host for the different target systems. For targets that were observed multiple times, the spectrum with the highest signal-to-noise ratio is shown.}
        \label{figure_contrast_spectra_mosaic}
\end{figure*}

\begin{figure*}
        \centering
        \includegraphics[width=0.98\textwidth]{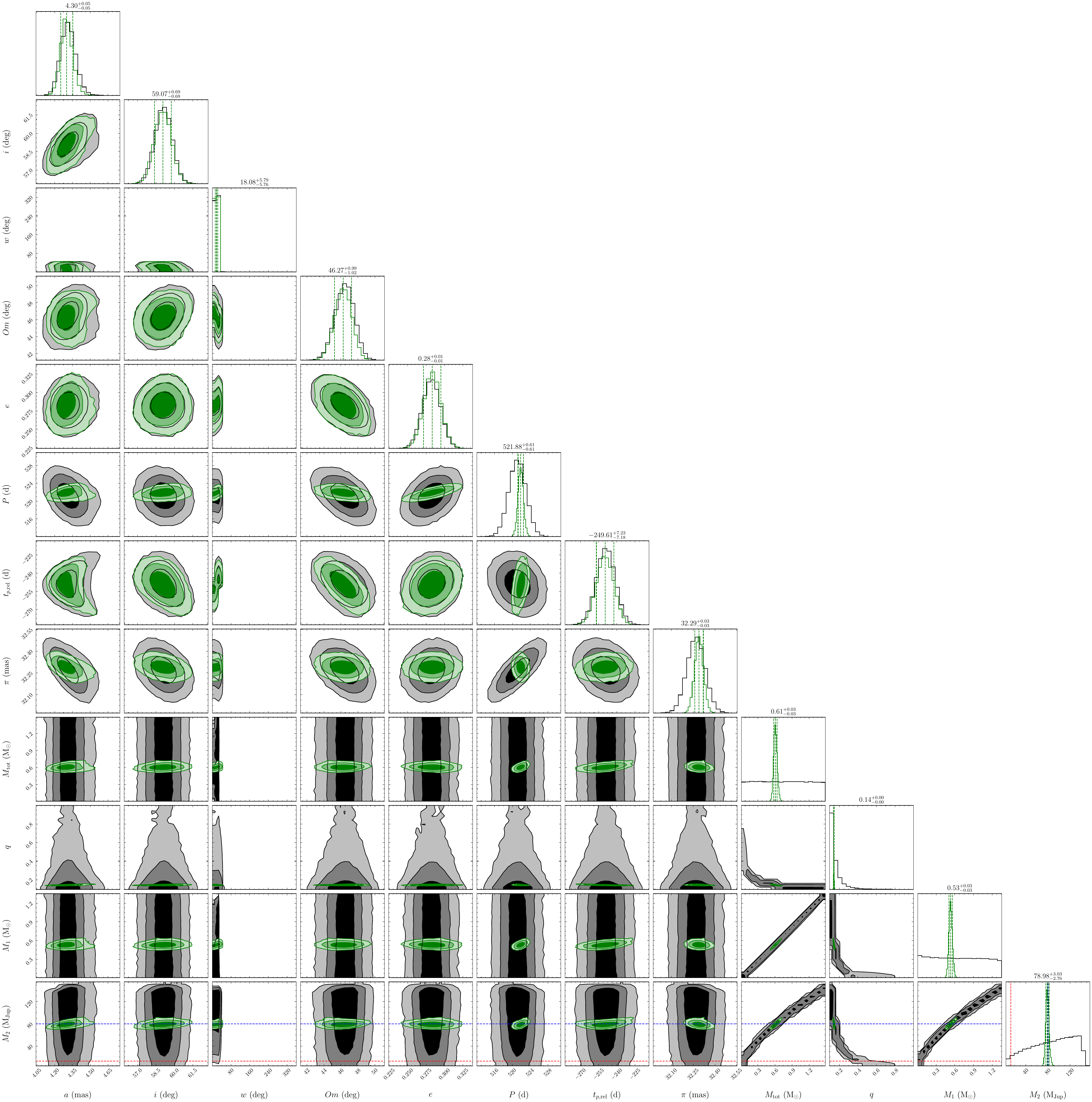}
        \caption{Corner plot showing the posterior samplings for the G ...6464 system. The \textit{Gaia}-only and the \textit{Gaia}-plus-GRAVITY run are presented in black and green, respectively. The indices 1 denote that the shown parameters describe the host star's orbit around the respective system's COM. The dashed blue and red lines in the $M_2$ row indicate the upper and lower BD mass boundaries, respectively.}
        \label{figure_corner_G_6464}
\end{figure*}

\begin{figure*}
        \centering
        \includegraphics[width=0.98\textwidth]{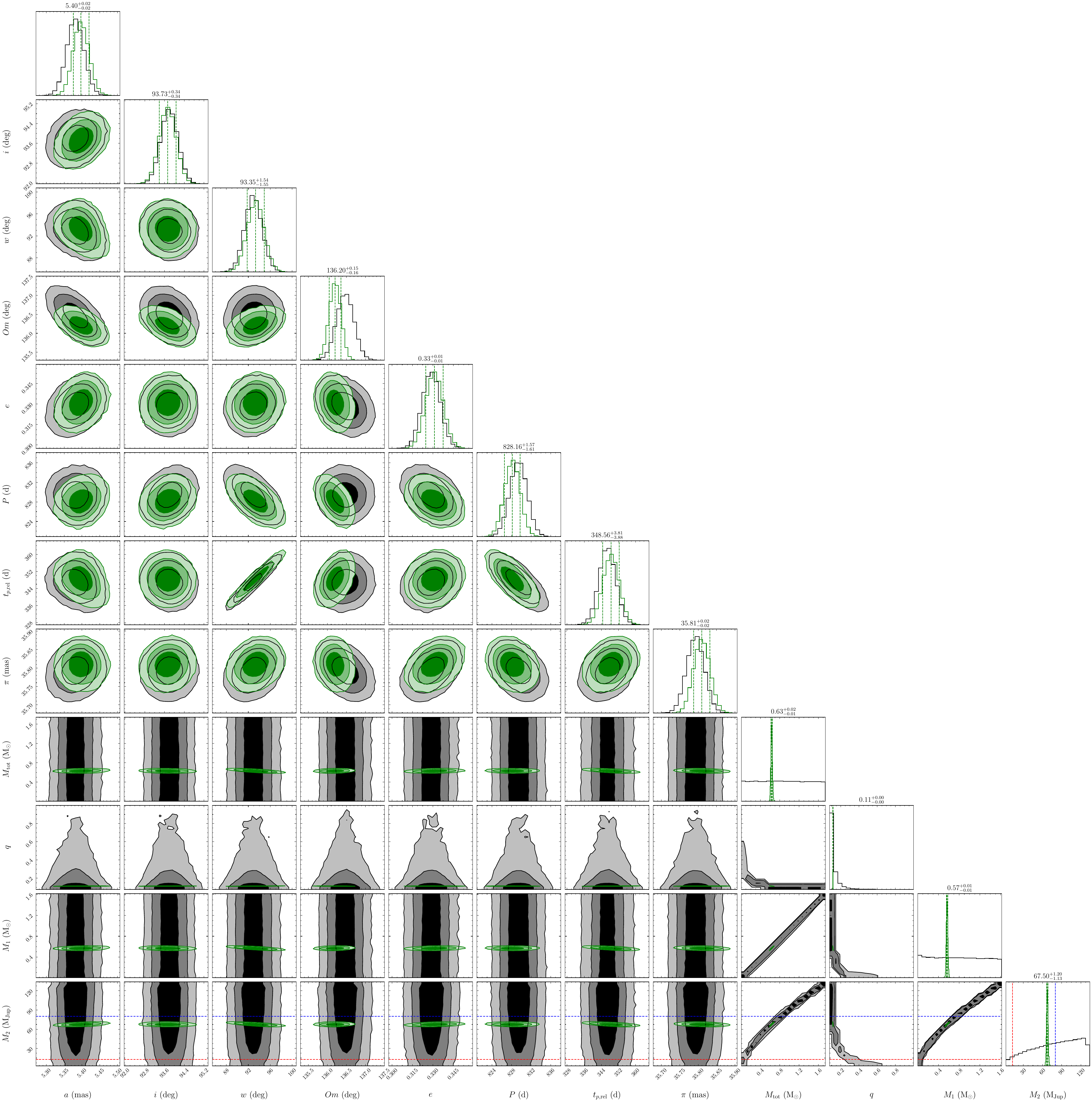}
        \caption{Same as Fig.~\ref{figure_corner_G_6464} but for the StKM1-1494 system.}
        \label{figure_corner_StKM1_1494}
\end{figure*}

Here, we present additional relevant tables and plots. Table~\ref{table_obs_log} presents the observation log for every target observed with the GRAVITY instrument. Table~\ref{table_stellar_mag} lists the stellar magnitudes used for the spectral fits. Table~\ref{table_numerical_results} showcases the inferred numerical results for each orbital parameter, the host and companion masses as well as the characteristics inferred from evolutionary models for each target system. Table~\ref{table_synth_comp_mags} lists synthetic companion magnitudes resulting from the application of the inferred ages and masses to evolutionary models. Figure~\ref{figure_contrast_spectra_mosaic} shows the companion-to-host K band contrast spectra measured by GRAVITY. The full corner plots of the respective MCMC runs for each target system are presented in Fig.~\ref{figure_corner_G_6464} to \ref{figure_corner_StKM1_1494}.

\begin{table*}[h]
    \centering
    \tiny
    \caption{Observation log of the GRAVITY follow-up.}
    \label{table_obs_log}
    \resizebox{\textwidth}{!}{%
    \begin{tabular}{lcccccccc}
    \toprule
    Target Name & Date & Start time (UT) & End time (UT) & $N_\mathrm{exp}$ / $N_\mathrm{DIT}$ / DIT (s) & airmass & $\tau_0$ (ms) & seeing & $(\Delta$RA, $\Delta$Dec$)_\mathrm{Fibre}$ (mas, mas) \\ 
    \midrule
    \midrule
    G ...6464 & 2022-11-09 & 01:56:47 & 02:57:59 & 16 / 4 / 30 & 1.82 & 0.006 & 0.560 & $(28, 53)$ \\
    G ...0976 & 2022-11-10 & 01:49:26 & 02:22:20 & 8 / 4 / 30 & 1.05 & 0.006 & 0.685 & $(38, -60)$ \\
    CD-50 869 & 2022-07-22 & 08:55:04 & 09:56:35 & 8 / 32 / 10 & 1.24 & 0.002 & 1.288 & $(50, -25)$ \\
    HD 17155 & 2022-07-22 & 07:31:51 & 08:29:51 & 8 / 32 / 10 & 1.44 & 0.002 & 1.26 & $(55, -15)$ \\
     & 2022-08-19 & 08:42:32 & 09:38:51 & 8 / 32 / 10 & 1.08 & 0.013 & 0.586 & $(54, 7)$ \\
    WT 766 & 2022-08-18 & 04:34:07 & 05:00:46 & 4 / 12 / 30 & 1.12 & 0.003 & 1.23 & $(-70, -60)$ \\
    G ...3872 & 2023-07-03 & 08:47:12 & 09:56:39 & 8 / 4 / 100 & 1.40 & 0.005 & 0.625 & $(65, -55)$ \\
              & 2023-10-27 & 06:12:01 & 06:47:37 & 4 / 8 / 30 & 1.27 & 0.002 & 1.176 & $(59, -28)$ \\
    G ...1920 & 2023-07-31 & 09:17:12 & 10:30:54 & 8 / 4 / 100 & 1.14 & 0.003 & 0.749 & $(44, 73)$ \\
    StKM1-1494 & 2023-05-08 & 06:11:37 & 07:19:57 & 8 / 4 / 100 & 1.06 & 0.002 & 0.976 & $(-54, 58)$ \\
    \bottomrule
    \end{tabular}
    }
    \tablefoot{$N_\mathrm{exp}$ denotes the number of exposures. $N_\mathrm{DIT}$ is the number of detector integrations per exposure, while DIT is the detector integration time. The atmospheric coherence time during the observation is given by $\tau_0$. $(\Delta$RA, $\Delta$Dec$)_\mathrm{Fibre}$ describes the placement of the fibre centre relative to the host star.}
\end{table*}

\begin{table*}[h]
    \centering
    \tiny
    \caption{Stellar magnitudes in different wavelength bands used for the spectral fits performed in Sect.~\ref{section_inferring_ages_from_evo}.}
    \label{table_stellar_mag}
    \resizebox{\textwidth}{!}{%
    \begin{tabular}{lccccccccc}
    \toprule
    Target Name & GAIA3.G$^1$ & GAIA3.Grvs$^1$ & 2MASS.J$^2$ & 2MASS.H$^2$ & 2MASS.Ks$^2$ & WISE.W1$^3$ & WISE.W2$^3$ & WISE.W3$^3$ & WISE.W4$^3$ \\ 
    \midrule
    \midrule
    G ...6464 & \SI{11.687 (2)}{} & -- & \SI{9.20 (2)}{} & \SI{8.61 (6)}{} & \SI{8.30 (3)}{} & \SI{8.21 (2)}{} & \SI{8.11 (2)}{} & \SI{8.00 (2)}{} & \SI{7.9 (2)}{} \\
    G ...0976 & \SI{13.708 (3)}{} & -- & \SI{10.86 (2)}{} & \SI{10.35 (2)}{} & \SI{10.08 (2)}{} & \SI{9.90 (2)}{} & \SI{9.73 (2)}{} & \SI{9.56 (3)}{} & \SI{8.6 (3)}{} \\
    CD-50 869 & \SI{9.463 (3)}{} & -- & \SI{8.07 (2)}{} & \SI{7.68 (6)}{} & \SI{7.53 (2)}{} & \SI{7.39 (3)}{} & \SI{7.53 (2)}{} & \SI{7.486 (17)}{} & \SI{7.59 (10)}{} \\
    HD 17155 & \SI{8.709 (3)}{} & -- & \SI{7.14 (2)}{} & \SI{6.62 (3)}{} & \SI{6.49 (3)}{} & \SI{6.47 (8)}{} & \SI{6.46 (2)}{} & \SI{6.486 (15)}{} & \SI{6.45 (5)}{} \\
    WT 766 & \SI{11.985 (3)}{} & \SI{10.177 (8)}{} & \SI{9.12 (3)}{} & \SI{8.48 (3)}{} & \SI{8.19 (2)}{} & \SI{8.03 (2)}{} & \SI{7.83 (2)}{} & \SI{7.671 (18)}{} & \SI{7.45 (14)}{}  \\
    G ...3872 & \SI{11.364 (3)}{} & -- & \SI{9.01 (3)}{} & \SI{8.35 (4)}{} & \SI{8.14 (3)}{} & \SI{7.98 (3)}{} & \SI{7.97 (2)}{} & \SI{7.857 (18)}{} & \SI{7.63 (11)}{} \\
    G ...1920 & \SI{11.688 (3)}{} & -- & \SI{9.52 (2)}{} & \SI{8.85 (3)}{} & \SI{8.68 (2)}{} & \SI{8.54 (2)}{} & \SI{8.51 (2)}{} & \SI{8.40 (2)}{} & \SI{8.5 (2)}{} \\
    StKM1-1494 & \SI{10.535 (3)}{} & -- & \SI{8.30 (2)}{} & \SI{7.71 (3)}{} & \SI{7.45 (3)}{} & \SI{7.34 (3)}{} & \SI{7.34 (2)}{} & \SI{7.268 (18)}{} & \SI{7.17 (10)}{} \\
    \bottomrule
    \end{tabular}
    }
    \tablebib{(1)~\citet{Gaia_DR3}; (2) \citet{2MASS_catalogue}; (3) \citet{ALLWISE_DR_Cutri_2013}.}
\end{table*}

\begin{table*}[h]
\centering
\renewcommand{\arraystretch}{1.7}
\caption{Orbital and evolutionary model parameters of the respective target systems.}
\label{table_numerical_results}
\resizebox{\textwidth}{!}{%
\begin{tabular}{lcccccccc}
\toprule
&
G ...6464 &
G ...0976 &
CD-50 869 &
HD 17155 &
WT 766 &
G ...3872 &
G ...1920 &
StKM1-1494 \\
\midrule
\midrule
$a_1$ (\SI{}{mas}) & $4.30^{+0.05}_{-0.05}$ & -- & $4.85^{+0.18}_{-0.16}$ & $8.12^{+0.14}_{-0.14}$ & $17.66^{+0.14}_{-0.13}$ & $5.15^{+0.03}_{-0.03}$ & $4.54^{+0.03}_{-0.02}$ & $5.40^{+0.02}_{-0.02}$ \\ 
$i$ (\SI{}{deg}) & $59.0^{+0.7}_{-0.7}$ & -- & $92.2^{+0.2}_{-0.2}$ & $71.5^{+0.4}_{-0.4}$ & $89.87^{+0.11}_{-0.10}$ & $152.8^{+1.0}_{-1.0}$ & $99.2^{+0.3}_{-0.3}$ & $93.7^{+0.3}_{-0.3}$ \\ 
$\omega_1$ (\SI{}{deg}) & $18^{+6}_{-6}$ & -- & $72.6^{+0.8}_{-0.8}$ & $95.7^{+0.4}_{-0.4}$ & $132.0^{+0.6}_{-0.6}$ & $348^{+3}_{-3}$ & $321.9^{+0.7}_{-0.7}$ & $93.3^{+1.6}_{-1.5}$ \\ 
$\Omega_1$ (\SI{}{deg}) & $46.2^{+1.0}_{-1.0}$ & -- & $113.7^{+0.3}_{-0.3}$ & $0.5^{+0.4}_{-0.3}$ & $49.46^{+0.06}_{-0.06}$ & $85.6^{+1.9}_{-1.9}$ & $27.3^{+0.2}_{-0.2}$ & $136.20^{+0.15}_{-0.16}$ \\ 
e & $0.284^{+0.011}_{-0.012}$ & -- & $0.722^{+0.019}_{-0.018}$ & $0.770^{+0.009}_{-0.009}$ & $0.623^{+0.007}_{-0.007}$ & $0.315^{+0.006}_{-0.006}$ & $0.638^{+0.007}_{-0.007}$ & $0.331^{+0.006}_{-0.006}$ \\ 
$P$ (\SI{}{d}) & $521.9^{+0.6}_{-0.6}$ & -- & $1780^{+80}_{-70}$ & $1463^{+12}_{-11}$ & $596.6^{+0.7}_{-0.7}$ & $983.1^{+0.9}_{-0.9}$ & $788^{+2}_{-2}$ & $828.2^{+1.6}_{-1.6}$ \\ 
$t_\mathrm{p, rel}$ (\SI{}{d}) & $-249^{+7}_{-7}$ & -- & $53^{+8}_{-9}$ & $360^{+8}_{-8}$ & $230.7^{+0.9}_{-1.0}$ & $398^{+3}_{-3}$ & $-71.8^{+1.3}_{-1.3}$ & $348^{+4}_{-4}$ \\ 
$\pi$ (\SI{}{mas}) & $32.29^{+0.03}_{-0.03}$ & -- & $18.877^{+0.011}_{-0.011}$ & $35.305^{+0.013}_{-0.013}$ & $74.92^{+0.04}_{-0.04}$ & $29.0560^{+0.018}_{-0.018}$ & $21.860^{+0.012}_{-0.013}$ & $35.81^{+0.02}_{-0.02}$ \\ 
$M_{tot}$ (\SI{}{M_\odot}) & $0.60^{+0.03}_{-0.03}$ & -- & $1.1^{+0.3}_{-0.2}$ & $0.77^{+0.05}_{-0.05}$ & $0.285^{+0.005}_{-0.005}$ & $0.653^{+0.011}_{-0.011}$ & $0.687^{+0.014}_{-0.013}$ & $0.632^{+0.016}_{-0.015}$ \\ 
$q$ & $0.142^{+0.003}_{-0.003}$ & -- & $0.094^{+0.008}_{-0.009}$ & $0.111^{+0.003}_{-0.003}$ & $0.349^{+0.003}_{-0.003}$ & $0.1182^{+0.0013}_{-0.0013}$ & $0.1642^{+0.0018}_{-0.0018}$ & $0.1135^{+0.0011}_{-0.0010}$ \\ 
$M_1$ (\SI{}{M_\odot}) & $0.53^{+0.03}_{-0.03}$ & -- & $1.03^{+0.3}_{-0.19}$ & $0.69^{+0.05}_{-0.04}$ & $0.211^{+0.004}_{-0.004}$ & $0.584^{+0.011}_{-0.010}$ & $0.590^{+0.013}_{-0.012}$ & $0.568^{+0.014}_{-0.014}$ \\ 
$M_2$ (\SI{}{M_{jup}}) & $79^{+3}_{-3}$ & -- & $102^{+19}_{-12}$ & $80^{+4}_{-4}$ & $77.3^{+1.4}_{-1.3}$ & $72.3^{+0.7}_{-0.7}$ & $101.5^{+1.3}_{-1.3}$ & $67.5^{+1.2}_{-1.2}$ \\

$\mathrm{Age}^\texttt{ATMO}$ (\SI{}{Myr}) & -- & -- & -- & -- & $670^{+40}_{-40}$ & $1950^{+170}_{-120}$ & -- & $670^{+40}_{-30}$      \\
$\mathrm{Age}^\mathrm{Baraffe}$ (\SI{}{Myr}) & -- & -- & -- & $760^{+260}_{-130}$ & $650^{+50}_{-40}$ & -- & -- & $630^{+40}_{-30}$      \\
$T_\mathrm{eff}^\texttt{ATMO}$ (\SI{}{K}) & -- & --  & -- & -- & $2328^{+5}_{-7}$ & $1750^{+8}_{-9}$ & -- & $2083^{+15}_{-12}$      \\
$T_\mathrm{eff}^\mathrm{Baraffe}$ (\SI{}{K}) & -- & -- & -- & $2319^{+11}_{-30}$ & $2316^{+17}_{-9}$ & -- & -- & $2054^{+15}_{-12}$      \\
$R^\texttt{ATMO}$ (\SI{}{R_{Jup}}) & -- & -- & -- & -- & $0.977^{+0.003}_{-0.004}$ & $0.844^{+0.003}_{-0.003}$ & -- & $0.941^{+0.004}_{-0.005}$      \\
$R^\mathrm{Baraffe}$ (\SI{}{R_{Jup}}) & -- & -- & -- & $0.988^{+0.011}_{-0.006}$ & $1.000^{+0.006}_{-0.007}$ & -- & -- & $0.967^{+0.004}_{-0.003}$      \\
\bottomrule
\end{tabular}
}
\tablefoot{The indices 1 denote that the shown parameters describe the host star's orbit around the respective system's COM.}
\end{table*}

\begin{table*}[h]
    \centering
    \renewcommand{\arraystretch}{1.7}
    \tiny
    \caption{Synthetic absolute magnitudes of the different companions.}
    \label{table_synth_comp_mags}
    \resizebox{\textwidth}{!}{%
    \begin{tabular}{lccccccccccc}
    \toprule
    Target Name & \texttt{MKO.J}$^1$ & \texttt{MKO.H}$^1$ & \texttt{MKO.K}$^1$ & \texttt{MKO.Lp}$^1$ & \texttt{MKO.Mp}$^1$ & \texttt{WISE.W1}$^2$ & \texttt{WISE.W2}$^2$ & \texttt{WISE.W3}$^2$ & \texttt{WISE.W4}$^2$ & \texttt{GAIA3.G}$^3$ & $(F_c/F_*)_\texttt{GAIA3.G}$ \\ 
    \midrule
    \midrule
    G ...6464 B  & -- & -- & -- & -- & -- & -- & -- & -- & -- & -- & --  \\
    G ...0976 B  & -- & -- & -- & -- & -- & -- & -- & -- & -- & -- & -- \\
    CD-50 869 B  & -- & -- & -- & -- & -- & -- & -- & -- & -- & -- & --  \\
    HD 17155 B   & -- & -- & -- & -- & -- & -- & -- & -- & -- & $16.0^{+7}_{-4}$ & \SI{0.014 (8)}{\percent} \\
    WT 766 B     & $11.45^{+0.09}_{-0.07}$ & $10.85^{+0.07}_{-0.08}$ & $10.55^{+0.08}_{-0.07}$ & $9.83^{+0.05}_{-0.06}$ & $10.44^{+0.09}_{-0.09}$ & $10.27^{+0.07}_{-0.05}$ & $10.21^{+0.08}_{-0.07}$ & $9.70^{+0.05}_{-0.05}$ & $9.69^{+0.05}_{-0.06}$ & $16.11^{+0.20}_{-0.15}$ & \SI{1.3 (2)}{\percent} \\
    G ...3872 B  & $12.88^{+0.11}_{-0.09}$ & $12.23^{+0.13}_{-0.11}$ & $12.24^{+0.18}_{-0.13}$ & $11.14^{+0.13}_{-0.11}$ & $11.55^{+0.06}_{-0.06}$ & $12.05^{+0.17}_{-0.16}$ & $11.29^{+0.07}_{-0.07}$ & $10.49^{+0.05}_{-0.05}$ & $10.38^{+0.05}_{-0.04}$ & -- & -- \\
    G ...1920 B  & -- & -- & -- & -- & -- & -- & -- & -- & -- & -- & -- \\
    StKM1-1494 B & $11.98^{+0.09}_{-0.09}$ & $11.33^{+0.08}_{-0.08}$ & $11.09^{+0.10}_{-0.10}$ & $10.19^{+0.08}_{-0.08}$ & $10.91^{+0.08}_{-0.07}$ & $10.74^{+0.11}_{-0.10}$ & $10.63^{+0.08}_{-0.07}$ & $9.96^{+0.05}_{-0.04}$ & $9.94^{+0.04}_{-0.04}$ & $17.28^{+0.20}_{-0.18}$ & \SI{0.026 (4)}{\percent} \\
    \bottomrule
    \end{tabular}
    }
    \tablebib{(1)~\citet{Tokunaga_MKO_filters} ; (2)~\citet{ALLWISE_DR_Cutri_2013} ; (3)~\citet{Gaia_DR3}.}
    \tablefoot{The \texttt{GAIA3.G} magnitude was computed via the \cite{Baraffe_Evolutionary_models_2015} model, while the \texttt{ATMO} model grid \cite{Phillips_ATMO} was utilised to calculate the other magnitudes. Where the companion mass lies outside of the mass range for which one or the other model is defined, only one model could be used. No magnitudes could be computed for targets lacking age (G ...6464 B, CD-50 869 B and G ...1920 B) or both age and dynamical mass (G ...0976 B). The rightmost column shows the companion-to-host flux ratio in the \texttt{GAIA3.G} band, which is of importance for the zero-companion-\textit{Gaia}-flux hypothesis.}
\end{table*}

\begin{figure*}
        \centering
        \includegraphics[width=0.98\textwidth]{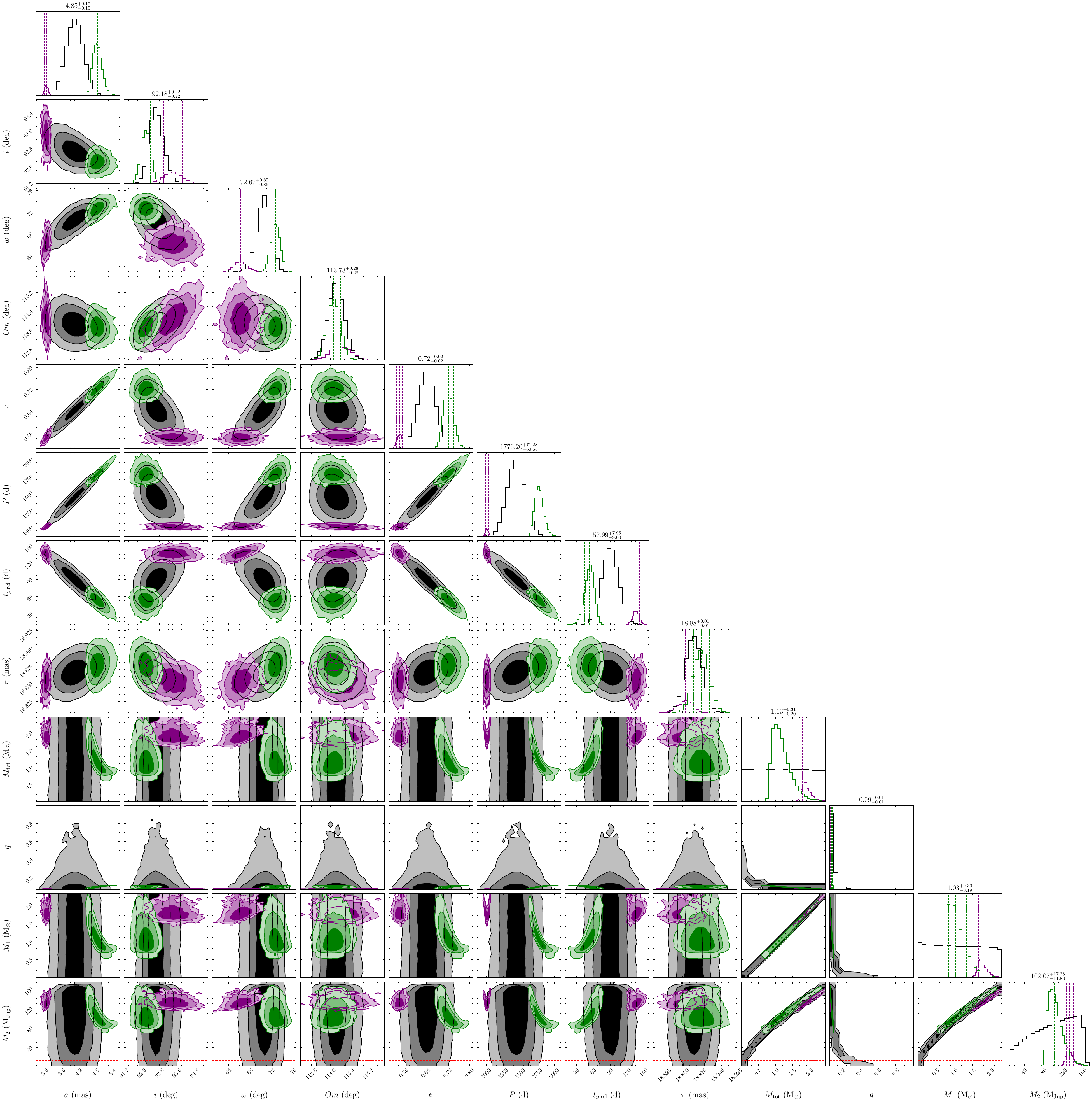}
        \caption{Same as Fig.~\ref{figure_corner_G_6464} but for the CD-50 869 system. This bimodal posterior sampling was separated into the preferred (green) and the secondary mode (purple).}
\end{figure*}

\begin{figure*}
        \centering
        \includegraphics[width=0.98\textwidth]{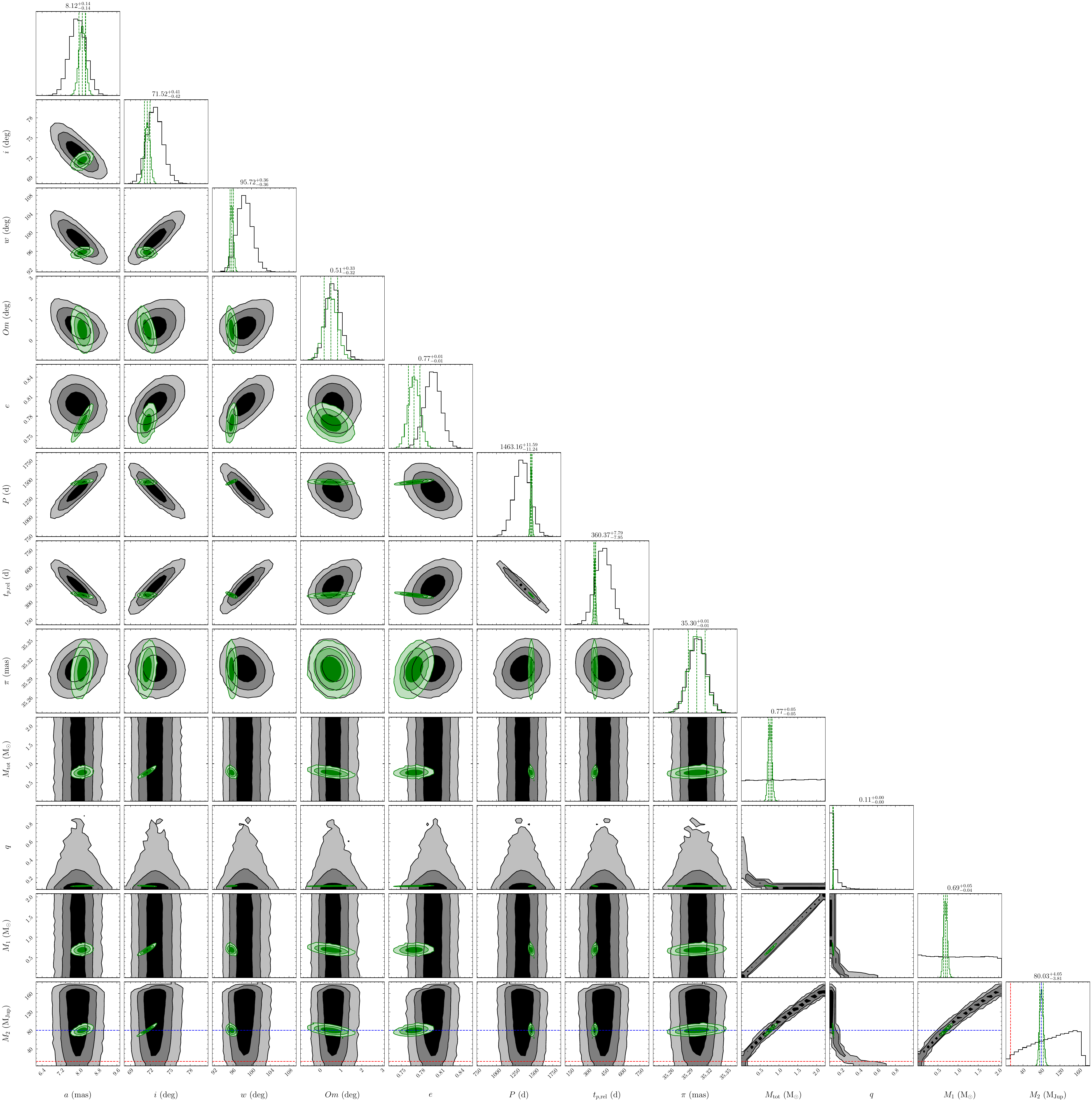}
        \caption{Same as Fig.~\ref{figure_corner_G_6464} but for the HD 17155 system.}
\end{figure*}

\begin{figure*}
        \centering
        \includegraphics[width=0.98\textwidth]{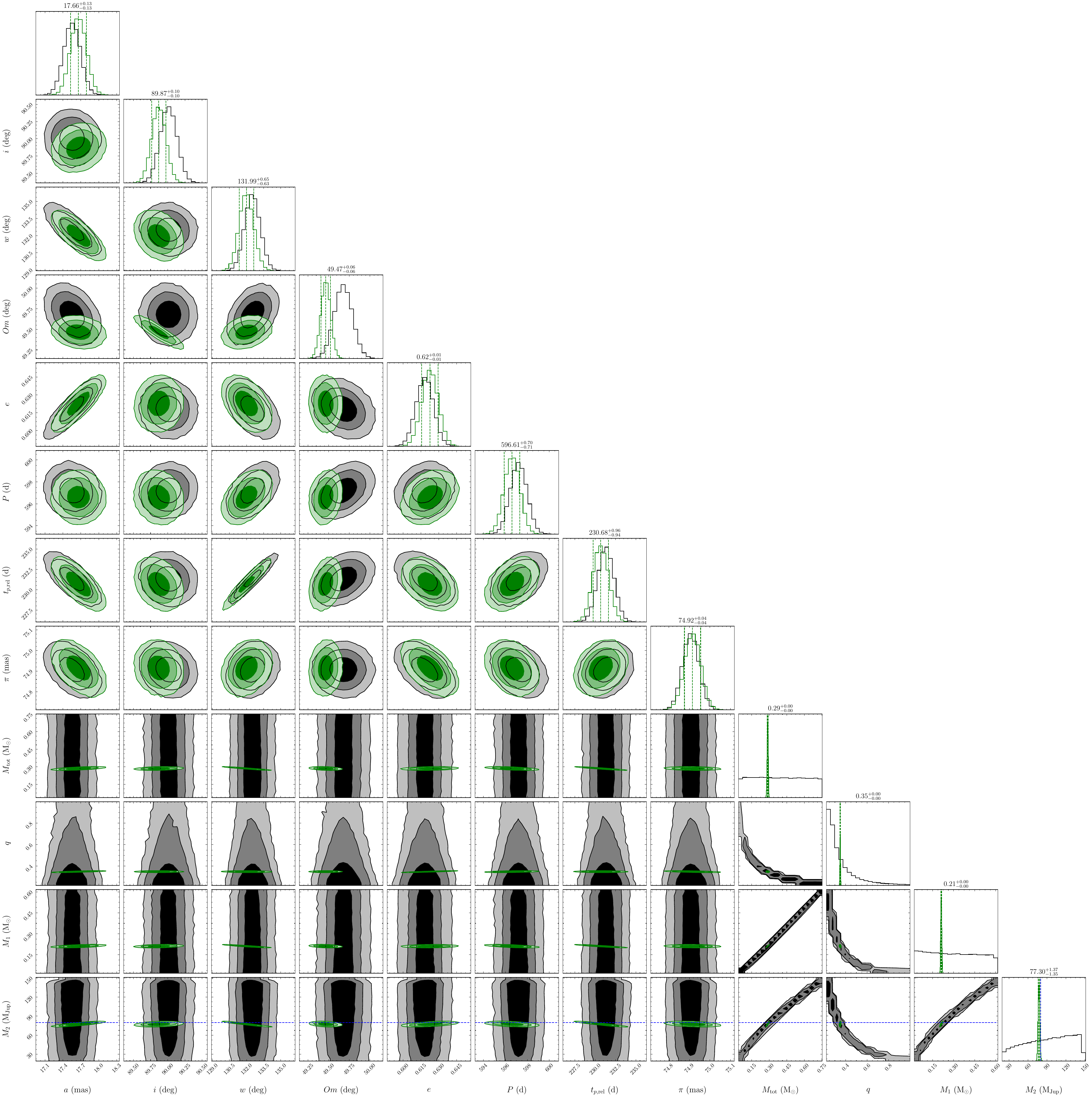}
        \caption{Same as Fig.~\ref{figure_corner_G_6464} but for the WT 766 system.}
\end{figure*}

\begin{figure*}
        \centering
        \includegraphics[width=0.98\textwidth]{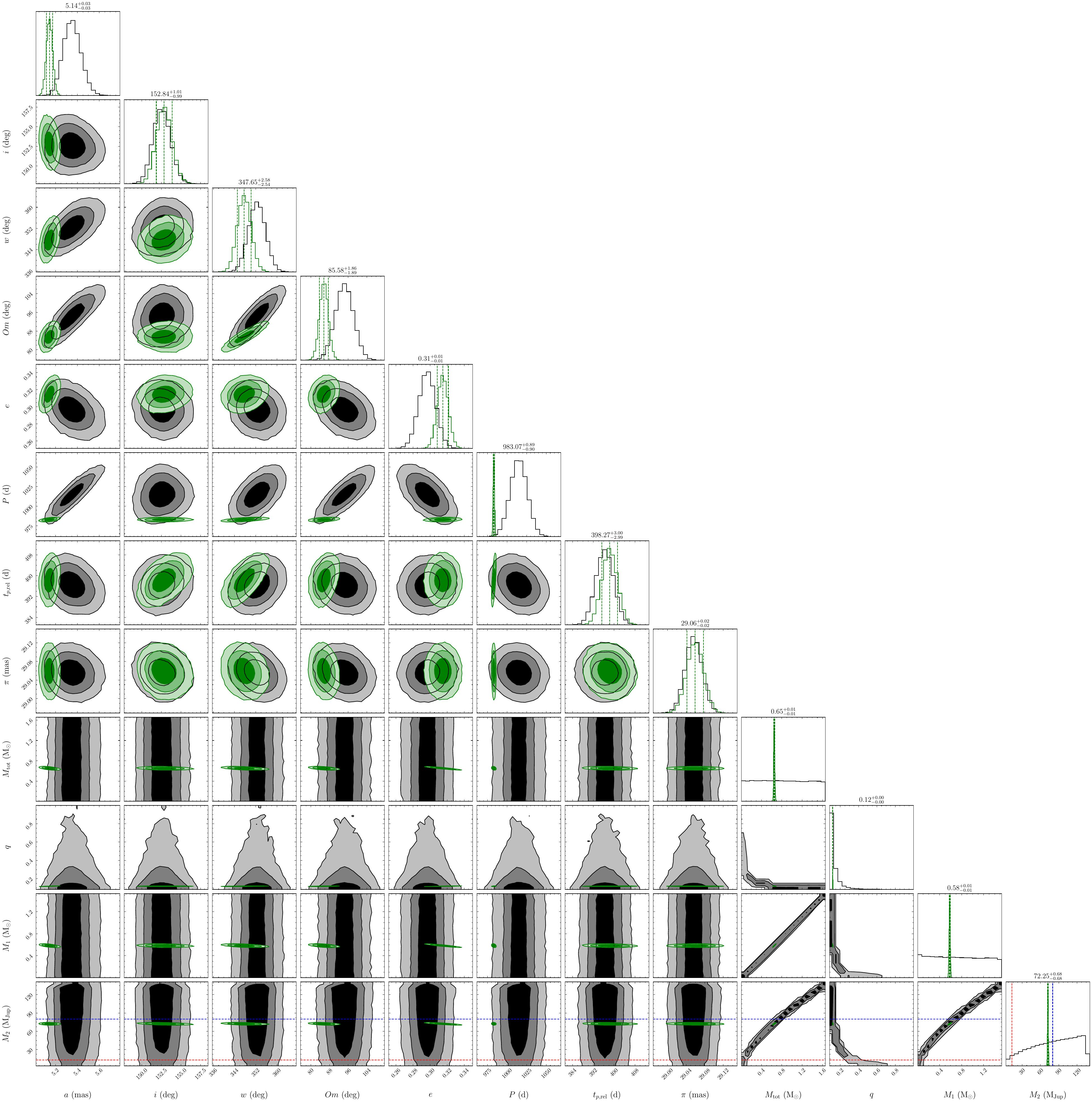}
        \caption{Same as Fig.~\ref{figure_corner_G_6464} but for the G ...3872 system.}
\end{figure*}

\begin{figure*}
        \centering
        \includegraphics[width=0.98\textwidth]{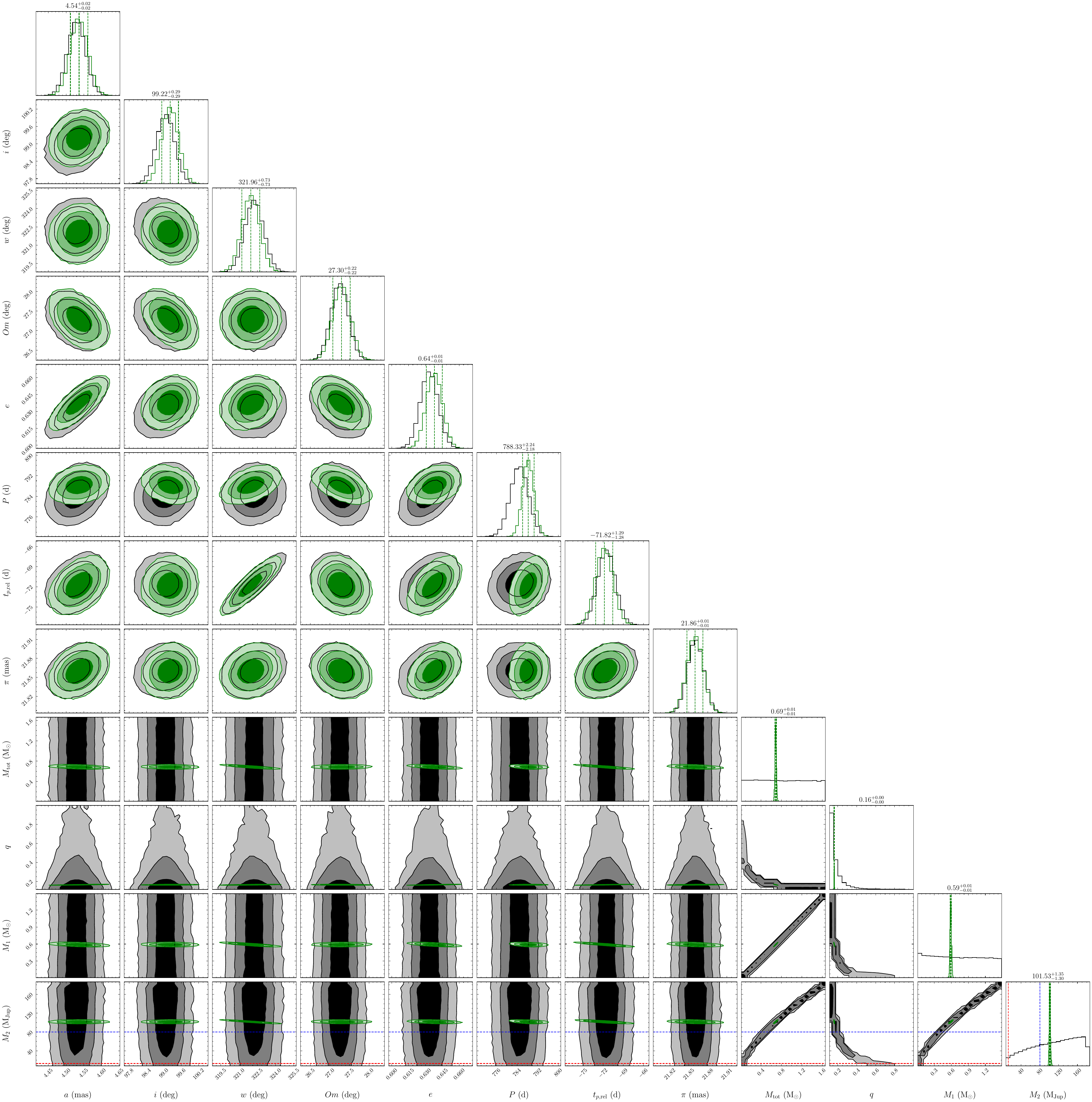}
        \caption{Same as Fig.~\ref{figure_corner_G_6464} but for the G ...1920 system.}
\end{figure*}

\end{appendix}

\end{document}